\documentclass[reprint,twocolumn,superscriptaddress,nofootinbib,aps,pre,floatfix,showkeys]{revtex4-1}

\usepackage[table]{xcolor}
\usepackage{graphicx,amssymb,verbatim,epsf,amsmath,setspace,float,dsfont, soul}
\usepackage{amsthm}
\usepackage{tabularx}
\usepackage{tikz}
\usetikzlibrary{matrix}
\usepackage{fancybox}
\usepackage{todonotes}

\newcommand{\R}{\mathbb{R}}
\newcommand{\N}{\mathbb{N}}

\newcommand{\norm}[1]{\left|\left|#1\right|\right|}
\newcommand{\inprod}[2]{\left\langle{#1},{#2}\right\rangle}
\newcommand{\btheta}{\boldsymbol{\theta}}
\newcommand{\mbf}[1]{\mathbf{#1}}
\newtheorem*{example}{Example}

\begin{document}

\title{Tensor clustering with algebraic constraints gives interpretable
groups \\ of crosstalk mechanisms in breast cancer}
\author{Anna Seigal} \email[]{seigal@berkeley.edu, joint first author}
\affiliation{Department of Mathematics, University of California,
  Berkeley, CA 94702, USA}
\author{Mariano Beguerisse-D\'iaz} \email[]{beguerisse@maths.ox.ac.uk, joint first author}
\affiliation{Mathematical Institute, University of Oxford, Oxford
  OX2 6GG, UK} 
\author{Birgit Schoeberl}
\affiliation{Novartis Institutes for BioMedical Research, Cambridge, MA 02139, USA}
\author{Mario Niepel} \affiliation{Ribon Therapeutics, Lexington, MA 02421, USA}
\author{Heather A. Harrington} 
\email[]{harrington@maths.ox.ac.uk}
\affiliation{Mathematical Institute, University of Oxford, Oxford
  OX2 6GG, UK}
\date{\today}

\begin{abstract}

We introduce a tensor-based clustering method to extract sparse,
low-dimensional structure from high-dimensional, multi-indexed
datasets.  This framework is designed to enable detection of clusters
of data in the presence of structural requirements which we encode as
algebraic constraints in a linear program. Our clustering method is
general and can be tailored to a variety of applications in science
and industry. We illustrate our method on a collection of experiments
measuring the response of genetically diverse breast cancer cell lines
to an array of ligands.  Each experiment consists of a cell
line-ligand combination, and contains time-course measurements of the
early-signalling kinases MAPK and AKT at two different ligand dose
levels.  By imposing appropriate structural constraints and respecting
the multi-indexed structure of the data, the analysis of clusters can
be optimized for biological interpretation and therapeutic
understanding.  We then perform a systematic, large-scale exploration
of mechanistic models of MAPK-AKT crosstalk for each cluster.  This
analysis allows us to quantify the heterogeneity of breast cancer cell
subtypes, and leads to hypotheses about the signalling mechanisms that
mediate the response of the cell lines to ligands. \end{abstract}


\maketitle

Muti-dimensional datasets are prevalent across the sciences; their
ubiquity and importance will only continue to
grow~\cite{Kolda09tensordecompositions, HVBKMSM, ABK, SSAA}. The
ever-increasing sophistication of datasets requires the development of
methods that preserve multi-dimensional structures and exploit them,
whilst maintaining interpretability of results. In addition, clustering biological data
is far from a straightforward task. There are multiple challenges,
including choosing an appropriate method for the
data~\cite{Lebart1984}, handling high-dimensional
data~\cite{VerSteeg2014,Madeira2004} and, importantly, the
consideration of the biological context of the problem, which must be
done almost on a case-by-case basis~\cite{VonLuxburg2009}.

Amongst the wide variety of clustering methods, constrained clustering
is an active field of research~\cite{Basu2008, Dao2017, Celebi2014,
  Wang2014, Li2018}. The most common approaches incorporate pairwise
{\it must-link} and {\it cannot-link} constraints to indicate whether
two items must or must not be in the same cluster~\cite{Wagstaff2001,
  Davidson2007}.  Other methods set constraints on what the possible
clusters can be, rather than constraining the elements in a
cluster~\cite{Mueller2010}. In these cases, there is a large pool of
candidate clusters from which those that meet selection criteria can
be chosen.

In this work we introduce a versatile data clustering framework based
on tensors and algebra to analyze high-dimensional datasets. One key
feature of our method is that it can incorporate general,
application-specific constraints on the composition of clusters, and
is guaranteed to find optimal partitions. The flexibility of the
method allows it to be used directly on a dataset (i.e., as a
standalone clustering tool), or in combination with other clustering
methods.

\begin{figure}[tp]
\includegraphics[width=.5\textwidth]{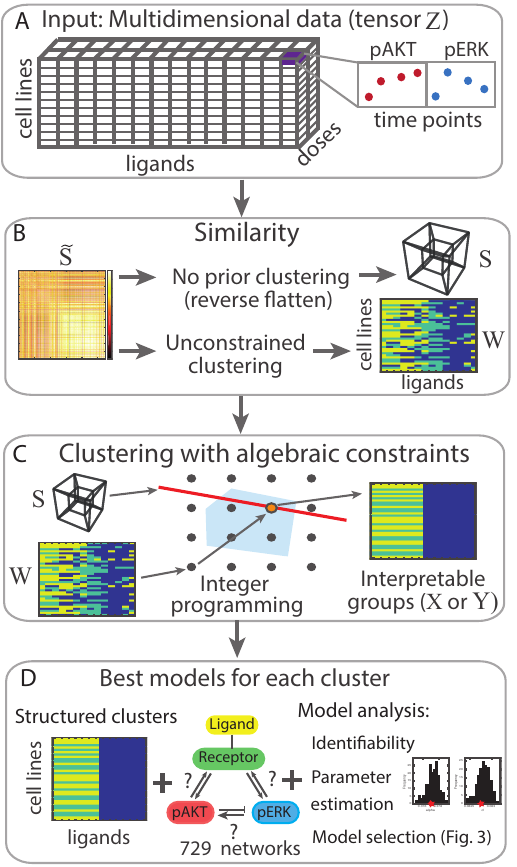}
\caption{Schematic of constrained tensor clustering method and model
  identification. (A) The complete set of experiments can be
  represented by the multi-indexed tensor $\mbf{Z}$; see Data
  section. (B) The similarity scores between experiments (each cell
  line/ligand combination) can be stored in a similarity matrix
  $\widetilde{\mbf{S}}$, that can be used to construct a similarity
  tensor $\mbf{S}$, or to find a preliminary clustering of the data
  $\mbf{W}$ that may not comply with the constraints. (C) Structured
  clustering via integer programming. The starting point can either be
  the similarity tensor $\mbf{S}$ or the pre-existing clustering
  $\mbf{W}$. The possible clusterings are represented by points on the
  grid. The red line is the value of the objective function
  (equations 6 and 7). The best integer value (orange point)
  is found inside the convex feasible region (blue).  (D) A large scale
  search for mechanistic models for each cluster involves
  parametrising, and ranking the best ODE models for each cluster.}
\label{fig:panel}
\end{figure}

We showcase our clustering framework on an extensive set of
time-course measurements of the activation levels of the
mitogen-activated protein kinase (MAPK) and phosphoinositide 3-kinase
(PI3K) pathways that are involved in cellular decisions and
fates~\cite{kriegsheim:2009,Marshall1995,Purvis:2013dd,Chen239}, and
are known to dysfunction in
cancer~\cite{Won01062012,JCP:JCP22647,Serra,Hanahan2011646,Baselga1175}.
The key signaling proteins and subtype responses in breast cancer
cells are known; however, among genetically diverse cell lines the
specific dysfunction mechanisms vary and are not well
understood~\cite{Niepel2014,Kolch2015,Heiser:2012dx}.  We examine a
set of experimental data \cite{Niepel2014} containing the response of
36 breast cancer cell lines after exposure to 14 ligands (growth
factors/signaling molecules). Each experiment measures the temporal
phosphorylation response of one cell line to one ligand.  Because the
dataset is {\it complete} (i.e., there is a measurement for every
combination of times, proteins, cell lines, ligands, and doses), we
can represent it as a tensor in five dimensions
(Fig.~\ref{fig:panel}A).

\begin{figure}[tp]
\centerline{\includegraphics[width=.4\textwidth]{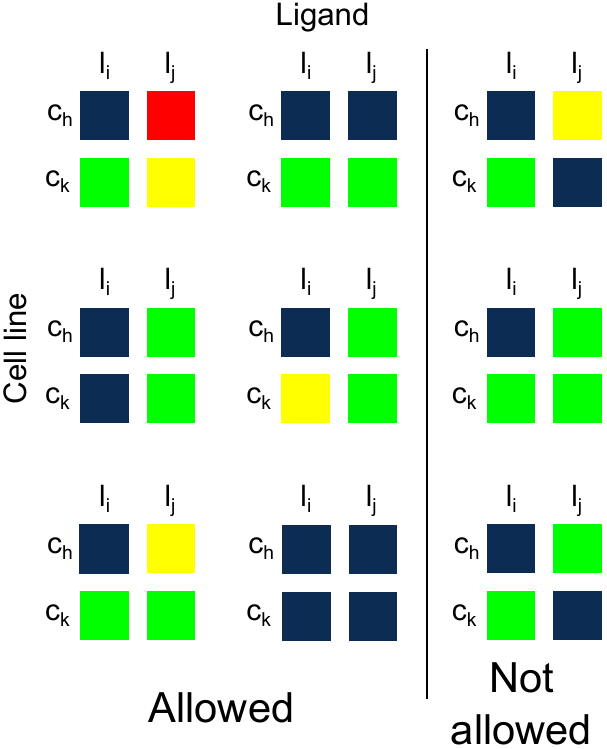}}
\caption{Examples of cluster shapes that are allowed and not allowed
  in our analysis of breast cancer data. The clusters on the first two
  columns are all rectangular, and thus allowed under our
  interpretability framework. The third column contains examples of
  non-rectangular clusters that are not allowed in our framework. Note
  that $j$ is not necessarily equal to $i+1$, and $k$ is not
  necessarily $h+1$.}
\label{fig:constraints}
\end{figure}

We find clusters of experiments subject to {\em interpretability
  constraints} (Fig.~\ref{fig:panel}B,C).  Our objective is to attribute 
differences between clusters
 to differences in the
underlying signaling mechanisms, so the composition of
the clusters must facilitate mechanistic interpretation.  For example, the cell lines in a cluster could share a mutation, and the ligands are
those whose effect is altered by the mutation. 
For this
reason, we constrain the clusters to be rectangular, i.e. to match a subset of cell lines with a
subset of ligands (Fig.~\ref{fig:constraints}).  
The constraints help to rule out similarities between
experimental measurements that are incompatible with a mechanistic
interpretation. The interpretability constraints take the form of
algebraic inequalities.

We introduce a new notion of tensor similarity, which we employ to
find optimal clusterings. The global optimality of the partitions is
guaranteed by leveraging results from integer programming. One of the
strengths of this approach, is that it can incorporate a pre-existing
non-rectangular partition obtained with other methods (e.g.,
conventional agglomerative clustering, $k$-means, spectral methods,
community detection on graphs) and find the nearest optimal
rectangular clustering. The distance between partitions is given by
the number of experiments whose clustering assignment changes.  Hence
this method can be used in conjunction with any other state-of-the-art
method and preserve the features that are compatible with the
constraints. Moreover, using the method from an initial partition is
computationally advantageous. The partition into clusters can be visualized by
color coding the grid of experiments according to their cluster
assignment (Fig.~\ref{fig:panel}C). Each box on the grid represents
the cluster assignment of an entire vector (or even a tensor) of data.

Once we obtain an optimal partition of the data, the second stage of our analysis is to search for
mechanisms that can explain the behavior of the experiments in each
cluster. We perform a systematic search for nonlinear ordinary
differential equation (ODE) models that reproduce the key dynamical
features of the time series in each cluster (Fig.~\ref{fig:panel}D).
To this end,
we construct, parametrize, and rank models for each cluster from a
pool of 729 candidate models.

\section{Tensors and Algebra}
\label{sec:algtensors}

{\it Data tensor.} We represent a multi-indexed dataset (e.g., the
complete dataset in Fig. 1A) as a tensor $\mbf{Z}$ of order $h$ in the
real numbers with size $n_1\times \dotsc \times n_h$ (i.e.,
$\mbf{Z}\in\R^{n_1\times \dotsc \times n_h}$, where $n_i\in \N$, and
$i=1, \dotsc , h$).  When the data is complete, every entry of the
tensor is filled with a number.  A full treatment of tensors is
available in~\cite{Kolda09tensordecompositions} and references
therein. We introduce here the tensor theory required for our
analysis.

{\it Similarity tensors.} In a similarity matrix the entry $(i,j)$
records the pairwise similarity of the two items labelled by
unidimensional indices $i$ and $j$.  We now introduce the
high-dimensional generalization of a similarity matrix, which extends
this to multi-indexed data. Suppose we want to compute the similarity
of the data indexed by $\mbf{i}=(i_1, i_2)$ and indexed by
$\mbf{j}=(j_1, j_2)$:
\begin{equation}
  s_{\mbf{i}, \mbf{j}} = \mathrm{sim}\left(\mbf{Z}(i_1,i_2,:, \dotsc,
  :),\, \mbf{Z}(j_1,j_2,:, \dotsc, :)\right),
  \label{eq:ten_sim}
\end{equation}
where $i_1,j_1 \in \{1, \dots, n_1\}$ and $i_2,j_2 \in \{1, \dots,
n_2\}$. The similarity function $\mathrm{sim}:\R^{n_3\times
  \cdots \times n_h} \times \R^{n_3\times \cdots \times n_h} \to \R$
computes the similarity between the data indexed by $\mbf{i}$ and
$\mbf{j}$ (e.g., correlation or cosine similarity).  In general, for
data indexed by the first $d$ dimensions, we have the multi-indices
${\bf i} = (i_{1}, \ldots, i_{d})$ and ${\bf j} = (j_{1}, \ldots,
j_{d})$.  The dimensions of $\mbf{Z}$ can be re-ordered as needed.  We
can construct a {\it similarity tensor} $\mbf{S}$ of order $2d$.  The
shape of $\mbf{S}$ is determined by the chosen dimensions of the data:
$\mbf{S}\in \R^{n_{1}\times\dotsc \times n_{d}\times n_{1}
  \dotsc\times n_{d}}$.  The similarity tensor and the similarity
matrix are related by {\it flattening} the tensor as follows. The
original data tensor $\mbf{Z}$ can be flattened (re-shaped) into a
data matrix $\widetilde{\mbf{Z}}\in\R^{N_1 \times N_2}$, where $N_1
=\prod_{r=1}^dn_{r}$, and $N_2=\prod_{r = d+1}^h n_{r}$.  Each row of
$\widetilde{\mbf{Z}}$ is a $N_2$-dimensional vector that corresponds
to multi-index $\mbf{i}$, and the length $N_2$ is the product of the
dimensions of $\mbf{Z}$ that are {\it not} included in $\mbf{i}$.

The similarity matrix between the rows of $\widetilde{\mbf{Z}}$ is
$\widetilde{\mbf{S}}\in \R^{N_1 \times N_1}$, which is obtained by
flattening the similarity tensor, ${\mbf S}$.  We summarize this
relationship in the following diagram:
\begin{equation*}
  \begin{tikzpicture}
    \matrix (m) [matrix of math nodes, row sep=3em, column sep=8em,
    minimum width=2em] 
    { \mathbf{Z} &  \mathbf{S}. \\
      \widetilde{\mathbf{Z}} &  \widetilde{\mathbf{S}} \\};
    \path[-stealth]
    (m-1-1) edge node [left] {\tiny flatten} (m-2-1) 
    edge node [above] {\tiny similarity of $\mbf{i}$ and  $\mbf{j}$} (m-1-2)
    (m-2-1) edge node [above] {\tiny similarity of rows} (m-2-2) 
    (m-2-2) edge node [right] {\tiny reverse flatten} (m-1-2) ;
  \end{tikzpicture}
\end{equation*}
To compute the similarity tensor $\mbf{S}$, we can simply flatten the
data tensor $\mbf{Z}$ into $\widetilde{\mbf{Z}}$, construct a
similarity matrix $\widetilde{\mbf{S}}$, and then reverse flatten it
into the desired $\mbf{S}$.  Note that $\mbf{Z}$ and
$\mbf{S}$ have the same number of entries as $\widetilde{\mbf{Z}}$ and
$\widetilde{\mbf{S}}$ respectively.
\begin{example}
  Let $\mbf{Z}\in \R^{10 \times 5 \times 3}$ be a tensor of order
  3. If $\mbf{i}=(i_1, i_2)$ is the multi-index, then $d=2$, $N_1
  =10\cdot 5=50$, and $N_2 =3$. The (order 4) similarity tensor
  $\mbf{S}$ has size $10\times5 \times 10\times 5$.  The similarity
  matrix $\widetilde{\mbf{S}}$ has size $50 \times 50$.  The flattened
  data matrix $\widetilde{\mbf{Z}}$ has size $50 \times 3$.
\end{example}

{\em Algebraic interpretability condition.} When clustering a set of
data points we typically seek to find a partition such that the points
within a cluster are more similar (or close) to each other than to the
rest of the data~\cite{Lebart1984}. In the simplest cases, there are
few restrictions on the clusters other than that the similarity or distance
be reflected in the cluster assignments.  In certain cases, imposing
restrictions on the clusters can be desirable or even
required~\cite{Celebi2014}.  Here we pursue {\it structured
  clustering}; that is, we impose restrictions on the shape of the
clusters in the tensor. In this application, we seek clusters with a
rectangular shape which allows us to interpret clusters in terms of
data-generating mechanisms (i.e., grouping cell lines/ligand
combinations to ensure mechanistic interpretation). We describe the
biological motivation for these constraints in the results section
(Sec.~\ref{sec:constraints}) and the mathematical details of the
method here.

A hard partition of a dataset represented as a tensor $\mbf{Z}$ of
size $n_1 \times \cdots \times n_h$ into $m$ clusters can be encoded
in two ways:
\begin{enumerate}
\item An $( n_1 \times \ldots \times n_d ) \times (n_1 \times \ldots
  \times n_d )$ tensor ${\bf X}$ in which the data have
  multi-indices ${\bf i} = (i_1, \ldots, i_d)$ and ${\bf j} = (j_1 ,
  \ldots, j_d)$, and:
  {\small
  \begin{equation}
    x_{\bf i j} =
    \begin{cases}
      0 & \text{ if ${\bf i}$ and ${ \bf j}$
        belong to the same cluster, } \\ 1 & \text{otherwise.}
    \end{cases}
    \label{eq:X}
  \end{equation}
  }
  The tensor ${\bf X}$ can be seen as a Boolean approximation of the
  distances between pairs of data points: $x_{\bf i j}=0$ if ${\bf i}$
  and ${\bf j}$ are `close' (in the same cluster), and $x_{\bf i j}=1$
  if they are `far' (in different clusters).  To ensure that ${\bf X}$
  encodes a valid clustering of the data, the three conditions of an
  equivalence relation must be met. These conditions are given by
  the following algebraic equations and inequality:
\begin{align}
  \text{Reflexivity:}& \quad x_{\bf i i} = 0, \nonumber \\ 
  \text{Symmetry:} & \quad x_{\bf i j} = x_{\bf j i }, \label{eq:const_X}\\ 
  \text{Transitivity:}& \quad 0
  \leq -x_{\bf i k} + x_{\bf i j } + x_{\bf j k} \leq 2. \nonumber
\end{align}
\item In a $n_1 \times \ldots \times n_d \times m$ tensor ${\bf Y}$,
  where
  {\small
  \begin{equation}
    y_{{\bf i} k} =
    \begin{cases} 1 & \text{ if the data indexed by
        ${\bf i}$ belongs to cluster $k$,} \\ 0 & \text{
        otherwise.}
    \end{cases}
    \label{eq:Y}
  \end{equation}
  }
  We require that $\sum_{k = 1}^m y_{{\bf i} k} = 1$, to ensure that
  each data item has been assigned to exactly one cluster.
\end{enumerate}
The tensors $\mbf{X}$ and $\mbf{Y}$ are related by the following equation:
$$ 1- {x}_{{\bf i}, {\bf j}} = \sum_{k = 1}^m {y}_{{\bf i},k} y_{{\bf j},k} .$$

{\em Integer Optimization.}  The structural or interpretability
conditions we have imposed on the clusters take the form of linear
constraints. These constraints, along with the fact that the tensors
are Boolean, allow us to find optimal tensors ${\bf X}$ and ${\bf Y}$
by solving an integer linear program~\cite{NW, BW}. Specifically, we
use the branch and cut algorithm~\cite{Mitchell} as we describe in the
Structured clustering section below.

\section{Data}

We examine an extensive experimental dataset detailing the temporal
phosphorylation response of signaling molecules in genetically diverse
breast cancer cell lines in response to different growth
factors~\cite{Niepel2014}.  This dataset is complete and can be
represented by a tensor $\mbf{Z}$ of order 5 whose dimensions
correspond to 36 cell lines, 14 ligands, two doses, three time points, and two
proteins (pERK, pAKT) (see the SI Appendix for more details).  In this
work each experiment is a set of measurements (for all time points,
doses, and proteins) for each cell line/ligand combination ($36\cdot
14=504$ experiments). Our goal is to find sets of experiments with a
similar response; consequently, the data structures we require are the
following:
\begin{align*}
& \mbf{Z}\in \R^{36 \times 14 \times 2\times 3 \times 2}, & & \text{(data tensor)}  \\
& \widetilde{\mbf{Z}}\in \R^{504\times 12},  & & \text{(flattened data tensor)}\\
& \mbf{S}\in \R^{36\times 14 \times 36 \times 14}, & & \text{(similarity tensor)} \\
& \widetilde{\mbf{S}}\in \R^{504\times504}. & &  \text{(similarity matrix)}
\end{align*}
Each experiment has a multi-index $\mbf{i}=(i_1, i_2)$, where
$i_1\in\{1,\dotsc,36\}$ and $i_2\in \{1,\dotsc, 14\}$.  We compute the
$504\times504$ cosine similarity matrix $\widetilde{\mbf{S}}$ of the
normalized rows of $\widetilde{\mbf{Z}}$ (see SI
  Appendix, II.B and II.C).

\section{Structured clustering}

Given a similarity tensor $\mbf{S}$, we seek the best partition of
experiments subject to the interpretability constraints: clusters be
rectangular with respect to cell lines and ligands (see
eq.~\eqref{eq:interp_cond} and results section).  This approach is
similar to those in Ref.~\cite{Madeira2004}; however, we do not
require the rectangles to be connected. This is because we do not
require a fixed order for the rows and columns of the data. This is an
important strength of our method: an ordering of the data is
artificial, and we seek clustering results that are not biased by
order.

We present two implementations of our method.  The first one does not
require previous knowledge about the clustering assignment of the
experiments, and provides an optimal clustering directly from the
similarity data. However, owing to the high
  computational costs of performing integer programming, this variant
  of our method is only appropriate for small datasets.  The
  computations can be sped up by employing heuristics for the
  integer optimization, see for example Ref.~\cite{ChenBatsonDang}.
  
  To tackle larger datasets, we present a second implementation that
begins with a pre-existing partition of the experiments into clusters
(not necessarily compliant with the constraints), which might
originate from {\it any} clustering method (e.g., using the reshaped
similarity tensor $\widetilde{\mbf{S}}$). This implementation then
reconstructs $\mbf{S}$ and finds the nearest optimal clustering
compliant with the constraints. Starting with an initial clustering
has the advantage that we can employ the best methods for clustering a
particular type of data, whose results we then refine to find clusters
that are compatible with the interpretability condition.  The initial
clustering must be chosen carefully to fit the application, and should
not be viewed as merely an initialization of the algorithm. Pairing
our method with a pre-existing clustering also has the advantage that
it significantly reduces computational cost (see SI Appendix, Figure
7).

\subsection{No prior clustering} 

When we do not have any prior clustering of the experiments, we work
directly on the similarity tensor $\mbf{S}$. The entries of this
tensor record the similarity of experiments $\mbf{i}$ and $\mbf{j}$,
where ${\bf i } = (i_1, i_2)$, ${\bf j} = (j_1, j_2)$, where the ranges
of indices are $i_1,j_1 \in \{ 1, \dotsc, 36\}$ and $i_2,j_2 \in \{ 1,
\dotsc, 14\}$.

The clustering assignments are recorded by the tensor $\mbf{X}$
defined in equation~\eqref{eq:X}. The rectangular-shaped
interpretability condition corresponds to three types of algebraic
constraints on the entries of $\mbf{X}$:
\begin{align}
  &  x_{i_1 i_2 j_1 j_2} = x_{i_1 j_2 j_1 i_2}, &    \nonumber \\ 
  &  0 \leq  x_{i_1 i_2 j_1 j_2} - x_{i_1 i_2 j_1 i_2} \leq 1, &
     \label{eq:interp_cond}\\
  & 0 \leq  x_{i_1 i_2 j_1 j_2}  - x_{i_1 i_2 i_1 j_2} \leq 1. & \nonumber  
\end{align}
We search over arrays $\mbf{X}$ that satisfy these conditions. The
experiments in the same cluster should have high similarity, so we
maximize the similarity between experiments in the same cluster. This
maximization is equivalent to solving the integer optimization problem
\begin{align}
  & \underset{\mbf{X}}\max & & \langle \mbf{S}, (\mathbf{1} -{\mbf X})
  \rangle + \lambda \langle \mathbf{1}, {\mbf X}
  \rangle, \label{eq:optimization} \\ & \text{subject to} & & b_l \leq
  \mbf{V}\cdot \mathrm{vec}(\mbf{X}) \leq b_u, \nonumber
\end{align}
where the tensors $\mbf{X}$ and $\mbf{S}$ are as above, $\langle
\cdot, \cdot \rangle$ denotes the entry-wise inner product, and
$\cdot$ represents matrix multiplication of the matrix $\mbf{V}$ by
the vector $ \mathrm{vec}(\mbf{X})$.  The $504^2\times 1$ vector $
\mathrm{vec}(\mbf{X})$ is the vectorized form of $\mbf{X}$, and
$\mathbf{1}$ is the tensor of ones with the same size as ${\bf
  X}$. The coefficient $\lambda$ is a regularization term introduced
to control the number of clusters. The matrix $\mbf{V}$ encodes the
constraints on $\mbf{X}$ given in eq.~\eqref{eq:const_X} and
eq.~\eqref{eq:interp_cond}. This matrix has over 1 million rows,
$504^2$ columns and is extremely sparse. The $k$th row of $\mbf{V}$
represents the $k$th constraint on the values of
$\mathrm{vec}(\mbf{X})$: the entry is the coefficient (which can be
$0$, $1$, or $-1$) with which each entry
of $\mathrm{vec}(\mbf{X})$ appears in the constraint. The $k$th entry
of $b_l$ and $b_u$ (which can be $0$, $1$, or $2$) give the lower and
upper bounds respectively of each linear inequality. We solve this
optimization program using the branch and cut
algorithm~\cite{Mitchell} via the IBM ILOG CPLEX Optimization
Studio~\cite{IBM2011}.

The resulting rectangular clusters are a sparse, low-rank
representation of the data. The tensor $\mbf{1}-\mbf{X},$ of size $(36
\times 14) \times (36 \times 14),$ gives a binary measure of the
distance between any two experiments. This tensor has sparse block
structure: it consists of $m$ cuboids of $1$s along the diagonal,
where $m$ is the number of clusters, and has zeros everywhere else. As
a consequence $\mbf{X}$ has low multilinear
rank~\cite{Lathauwer:2000}, bounded above by $(m,m,m,m)$, which is
less than the maximum possible value of $(36,14,36,14)$.

\subsection{Pre-existing clusters}

When we have a pre-existing or initial non-rectangular clustering of the
experiments, we find the nearest structured clusters using linear
integer optimization. The input to this method is an initial partition
of the 504 experiments into $m$ clusters. We then modify the cluster
assignments to reach the closest possible interpretable, structured
clustering.

 The initial clustering is encoded by a partition tensor, ${\mbf{W}}$,
 of size $36\times 14 \times m$
$$ w_{{\bf i} k} = \begin{cases} 1, & 
\text{ ${\bf i}$ is in cluster $k$,} \\ 0, & \text{ otherwise,} 
\end{cases}$$
where ${\bf i} = (i_1, i_2)$ indexes an experiment. The new clusters
are encoded by a tensor $\mbf{Y}$ of the same size (defined according
to equation~\eqref{eq:Y}). In order to have rectangular clusters, the
entries of $\mbf{Y}$ must satisfy the conditions
{\small
\begin{align*}
  \sum_{r=1}^m y_{ijr} = 1, &\quad \text{(unique cluster assignment)} \\
  -1 \leq y_{ikr} + y_{jlr} - y_{ilr} \leq 1. & \quad \text{(interpretability condition)}
\end{align*}
}
As before, we use the branch and cut algorithm to obtain the
global optimum (given $\mbf{W}$) for the optimization problem
\begin{align}  
\underset{\mbf{Y}}\max \quad \langle \mbf{W}, \mbf{Y} \rangle.
\label{eq:opt_Y}
\end{align} 
The inner product $ \langle \mbf{W} , \mbf{Y} \rangle$ sums the number
of clustering assignments unchanged by converting the initial
unstructured clustering into a clustering that satisfies the
interpretability constraints.

We obtain the tensor $\mbf{Y}$, of size $36 \times 14 \times m$ by
solving the optimization problem in eq.~\eqref{eq:opt_Y}. As with
$\mbf{X}$, the tensor $\mbf{Y}$ also has sparse and low-rank
structure. Its $m$ two-dimensional slices, each a matrix of size $36
\times 14$, have rank two and block structure with a rectangle
populated by $1$s and all other values equal to $0$.

\section{Results}
\label{sec:results}

\subsection{Biological interpretation of constraints}
\label{sec:constraints}

  Each experiment in our data is indexed by $(c_i,l_j)$,
where $c_i$ is the $i$th cell line, and $l_j$ is the $j$th ligand. A
high similarity between experiments suggests the possibility of a
common underlying biological mechanism. This is the basic notion that
underpins the constraints in our clustering method which force the
clusters to pair a subset of the cell lines with a subset of the
ligands in such a way that each cluster must be rectangular, 
although possibly disconnected (Fig.~\ref{fig:constraints}).  The
motivation behind this constraint is to enable the interpretation that
the experiments in each cluster are generated by the same biological
mechanism (e.g., if they share a feature such as a genetic mutation).
The difference between our constrained approach and conventional
clustering is that in the latter a high similarity is enough to
cluster two experiments together. In our approach similarity alone is
not enough, we also require that the observations admit the same
mechanistic interpretation.  For example, suppose that two experiments
$(c_h, l_i)$ and $(c_k, l_j)$ belong to the same cluster.  If we
swapped the ligands (i.e., we looked at the experiments in the
diagonally opposite entries $(c_k, l_i)$ and $(c_h, l_j)$), under the
assumption that the cell lines share the same signalling mechanism,
these experiments should also be in the same cluster because we expect
them to respond in a similar way (see left columns of
Fig.~\ref{fig:constraints}). If, however, $(c_h, l_i)$ and
$(c_k, l_j)$ are clustered together but $(c_k, l_i)$ and $(c_h, l_j)$
are not in the same cluster (see right column of
Fig.~\ref{fig:constraints}), it would be more difficult to assign
mechanistic interpretations to the clusters.

\subsection{Interpretable groups by mutation and receptor subtype}

\begin{figure*}[tp]
  \centerline{\includegraphics[width=.8\linewidth]{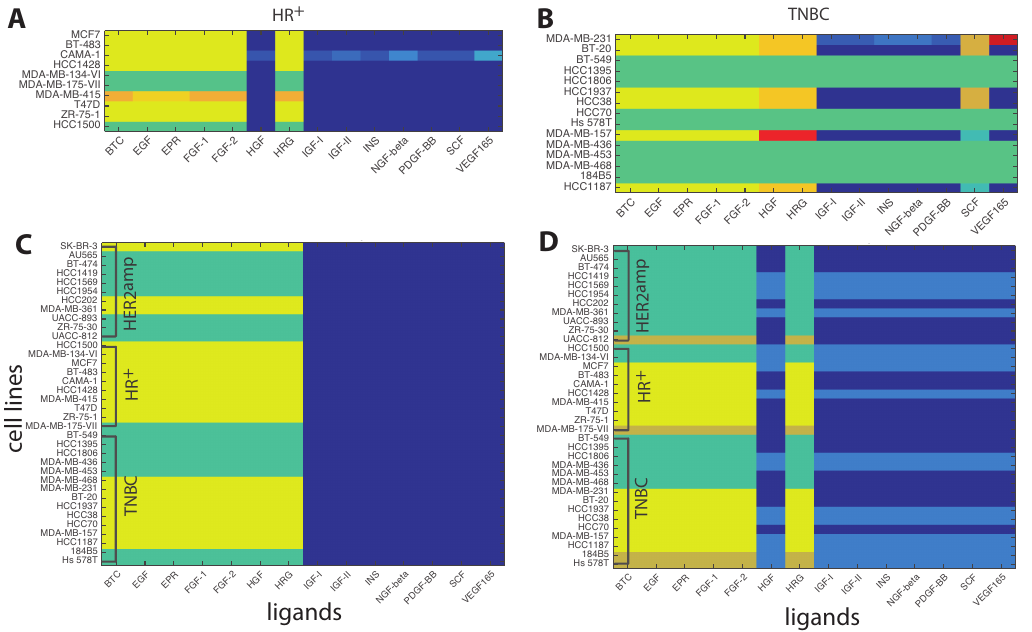}}
  \caption{ Tensor-based structured clustering. (A) TNBC clustering
    with no prior clustering information.  (B) HR$^{+}$ clustering
    with no prior information. (C) Clustering of all cell lines
    starting from an initial partition into three clusters. (D)
    Clustering from an initial partition into five clusters. {Note
      that the colours on the grid represent clustering assignments,
      and are not reflective of the intensity of any single
      parameter.}}
\label{fig:three-clusters}
\end{figure*}

In a clinical setting, prognosis and treatment decisions for breast
cancer are guided by tumor grade, stage and clinical
subtype~\cite{NCI}, which is based on the presence of cellular
receptors:
\vspace{-1em}
\begin{itemize}
\item HER2$^{\rm{amp}}$ cells are characterized by amplification of
  the HER2 gene, leading to over-expression of the ErbB2 receptor
  tyrosine kinase;
\item HR$^{+}$ cells are characterized by the expression of the
  estrogen receptor (ER) or progesterone receptor (PR);
\item Triple negative breast cancer (TNBC) cells are negative for
  HER2 amplification, and express ER and PR at low levels.
\end{itemize}
\vspace{-1em}
We compare the clusters from our method with the three standard
clinical subtypes above. We also compare our clusters with the
mutational status of the cell lines~\cite{Hollestelle:2007fb,COSMIC},
and with their drug response~\cite{hafner:2016},
and with the findings from the previous clustering and analysis of this data set, found in~\cite{Niepel2014}.

We first investigate a fine-grained classification within each of the
three clinical subtypes.  A summary statistic between 0 and 1 (based
on the cosine similarity, see SI Appendix II.B) quantifies the
within-class variation for each clinical subtype. A score of 0
indicates complete homogeneity, and 1 indicates complete
heterogeneity.  The HER2$^{\rm{amp}}$ cell lines show comparatively
little variation, with an average difference score of 0.086. The TNBC
and the HR$^{+}$ cell lines have an average difference score of 0.224
and 0.334.  We obtained clusters without prior knowledge of an initial
clustering by solving the optimization
problem~\eqref{eq:optimization}. The results (shown in
Fig.~\ref{fig:three-clusters}A,B) identify heterogeneity within each
subtype as well as cell lines of particular interest. 

Figure~\ref{fig:three-clusters}A shows the clustering of the HR$^{+}$
cell lines.  Cell line MDA-MB-415 stands out for its response to
so-called high-response ligands~\cite{Niepel2014} (ligands to the left
of HRG in Fig.~\ref{fig:three-clusters}B).  Among all cell lines,
MDA-MB-415 has the second highest susceptibility to the drugs
Ixabepilone, Methylglyoxal and PD~\cite{hafner:2016}.  CAMA-1 cell
line is distinctive in its response to the low-response ligands (to
the right of HRG), which might help explain why it is particularly
susceptible to both (Z)-4-Hydroxytamoxifen and TCS
PIM-11~\cite{hafner:2016}.  The TNBC cell lines are divided into 12
clusters (Fig.~\ref{fig:three-clusters}B), which mirror the
heterogeneous behavior of TNBC in the clinic~\cite{Podo2010209}.  All
but one TNBC cell lines with a PTEN mutation appear in the green
cluster. The only exception is the HCC1937 cell line, which has a PTEN
mutation but appears in the yellow cluster. The cluster assignment of
cell lines MDA-MB-231 and MDA-MB-157 is markedly different from that of the
other cells across the ligands.  These assignments might be explained
by the mutational status of the cell lines; MDA-MB-231 is the only
cell line with an NF2 mutation or a BRAF mutation, whereas MDA-MB-157
is the only cell line with an NF1 mutation. The bright orange cluster
contains five cell lines (all but HCC1937) with the same two CDKH2A
mutations.

  The HER2$^{\rm{amp}}$ cell lines cluster together for all
ligands except for the MDA-MB-361 cell line. This is the HER2$^{\rm
  amp}$ cell line most resistant to HER2-targeted therapy such as
Lapatinib~\cite{hafner:2016}. In fact, its resistance to Lapatinib
exceeds that of some TNBC cell lines (HCC2185 and MDA-MB-453).  The
grouping of the rest indicates the consistency among all other
HER2$^{\rm{amp}}$ cell lines (see SI Appendix II.B).

{\it Clustering all cell lines}. To cluster all cell lines we solve
the optimization problem~\eqref{eq:opt_Y} which requires an initial
``seed'' clustering of the experiments.  We obtained our initial
clustering by first constructing a graph of experiments from the
similarity matrix $\widetilde{\mbf{S}}$ using the Relaxed Minimum
Spanning Tree
algorithm~\cite{Vangelov2014,Beguerisse2013,Beguerisse2014}. Then we
used the Markov Stability community detection
method~\cite{Delvenne2010,Delvenne2013} to find robust partitions of
the experiments into three, five and seven groups (see SI Appendix,
Fig.~3).

From the initial partition into three clusters, we obtain three
rectangular clusters (Fig.~\ref{fig:three-clusters}C).  These groups
respect the broad division of the cell lines seen in
Fig.~\ref{fig:three-clusters}A,B, which is a sign of the consistency
between the two implementations of our method.  Of these, we find that
two groups of ligands correspond to previously reported high active
expression profiles (yellow and green) and one to muted profiles
(blue), respectively~\cite{Niepel2014}.  Within the more highly active
group, the HR$^{+}$ cell lines are predominantly in the yellow
cluster, while the HER2$^{\rm{amp}}$ cells are in the green
cluster. This separation of the HR$^{+}$ and HER2$^{\rm{amp}}$
clinical subtypes is entirely data driven and supports the notion that
our method is indeed able to find interpretable groups.  The cell
lines that are not clustered according to their subtype reflect
previous findings that neither growth factor responses nor sensitivity
to drugs that target signal transduction pathways is uniform within
clinical subtypes~\cite{Niepel2013,Niepel2014,Heiser:2012dx}. The TNBC
cell lines are divided between the yellow and green clusters,
providing further evidence of the heterogeneity in TNBC cell
lines~\cite{Kirouac2016,Kirouac2013,Kirouac2015,Shah2012,Shastry2013,Niepel2013}.

When we start from the initial non-rectangular clustering into five
groups, the resulting rectangular clusters split the ligands into a
low response group (blues) and high response (green, yellow,
brown). This split is nearly the same as we obtained before
(Fig.~\ref{fig:three-clusters}D). Note that the difference in the
ligand HGF may be due to the fact that it is not part of the ErbB nor
the FGF families of ligands.  The HER2$^{\rm{amp}}$ cell lines are now
all assigned to the green cluster and there are only three HR$^+$ cell
lines not assigned to the yellow cluster.  A new brown cluster
consists of cell lines: MDA-MB-175-VII (classified as a HR$^+$),
UACC-812 (HER2$^{\rm{amp}}$), 1845B5 (TNBC) and HS578T (TNBC).  While
none of them has the same cell classification or genetic mutation,
all cell lines in the brown cluster show high susceptibility to the drug
Gefitinib~\cite{hafner:2016}.  Note that MDA-MB-175-VII is the only
HR$^+$ cell line that is not assigned to the yellow group in either
three or five clusters; this might be due to the fact that this cell
line carries a unique chromosomal translocation. The translocation
leads to the fusion and amplification of neuregulin-1 which signals
through ErbB2/ErbB3 heterodimers~\cite{Schaefer:1997eu,Liu:1999fp},
and could be the underlying cause of the cell line's unique
sensitivity to ErbB-targeting drugs such as Lapatinib or
Afatinib~\cite{Niepel2013,Heiser:2012dx}.

We compare the results from our clustering method to the
  original analysis of this data set~\cite{Niepel2014}. In our
  analysis, we are able to obtain simultaneously meaningful subsets of
  {\it both indices} (the cell lines, and the ligands), without biases
  given to either index or to the ordering of the data. In contrast, in the unstructured
  clusters shown in~\cite[Figure 3]{Niepel2014}, the interpretation of
  the results required aggregating information to study how the
  effects vary with each cell line or with each ligand individually,
  but not simultaneously~\cite[Figure 4]{Niepel2014}. Our method allowed the clinical
  subtypes to be recovered from the data, based on temporal responses
  to a detected subset of the ligands. Exceptions to this
  classification provide biological hypotheses for possible subsequent
  investigation. In contrast, the clinical subtypes were not
  detectable from the clustering assignments of the temporal data made
  in~\cite[Figure 3]{Niepel2014}.

The clustering that begins from an initial partition into
seven groups shows high consistency with the five cluster case
(see SI Appendix, Figure 6). We therefore continue our
analysis on the five rectangular clusters.

\subsection{Systematic model identification}

We now analyze the response of the five structured groups found in the
previous section (Fig.~\ref{fig:three-clusters}D) to obtain a
mechanistic insight about about the cell line/ligand combinations in
each cluster.  We consider 729 possible
ODE network models, and then perform systematic model analysis of the 44 that are structurally
identifiable with the given data. We test structural identifiability, a prerequisite for performing parameter estimation and
model selection, using Daisy  \cite{DAISY}.
  Then, we parametrize, rank and choose
the models that best represent each cluster's response. As a result,
we have a list of candidate signaling mechanisms for each cluster
which provides more information than the statistical predictions of
the sensitivity of MAP Kinase drug targets (e.g., ErbB drug
class)~\cite{Niepel2013}.

\begin{figure}
\centerline{\includegraphics[width=.45\textwidth]{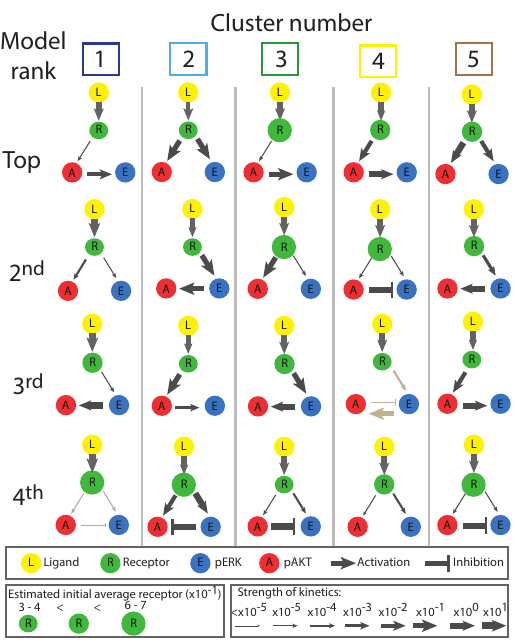}}
\caption{ The top four models for each cluster according to the AICc
  ranking. The strength of interactions are indicated by the size of
  the arrow. The grey arrows indicate a blocking mechanism for
  inhibition. Black inhibition arrows indicate a removal mechanism for
  inhibition. }
\label{fig:aic}
\end{figure}

Models of the MAPK and AKT pathways have been studied under a variety
of biological and modeling assumptions~\cite{Heinrich2002,Huang1996,
  Beguerisse2016}, including pathway
crosstalk~\cite{Fujita2010,Won01062012,Fey:2012ge,Chen239}.  Here we
consider simple models to ensure the parameters are at least locally
identifiable so there are a finite number of parameter values to fit
the data. 
We construct nonlinear
ordinary differential equation models to describe the dynamics of the
AKT and ERK signaling pathways. See the SI Appendix for a synopsis of MAPK models and details of their construction. Briefly, these models include three molecular
species: Receptor (R), pERK (E) and pAKT (A). Since the data contains
the response of pERK and pAKT, we assume that the receptor must
phosphorylate ERK and/or AKT. We consider positive, negative, or no
interaction between pERK and pAKT under different types of kinetic
regimes (mass-action or Michaelis--Menten) and different types of
inhibition (blocking/sequestration or removal/degradation). The
combination of these features results in the 44 structurally
identifiable models that we study in further detail. Each model
corresponds to a different mechanistic hypothesis of the dynamics in
the pathways (see SI Appendix III.C).  To find the models that best
describe the response of each of the five clusters, we estimate
parameters using the Squeeze-and-Breathe
algorithm~\cite{Beguerisse2012b}, and rank them using the Akaike
information criterion score (AICc) (see SI Appendix IV.D).  The best
models for each cluster are shown in Fig.~\ref{fig:aic}.

The AICc, used for model selection, penalizes more complex models;
therefore it is not surprising that the top models are the simplest
ones.  The best models for each cluster have different feedback
strengths (parameter values) and network topologies (see
Fig.~\ref{fig:aic}); this supports the hypothesis that mutations may
play a role in the dynamics. Although the values of the parameters
vary, the model with arrows from the receptor (R) to pAKT and pERK appears
in all clusters, which is in line with how cells are understood to
operate.  We remark that cluster~4, which corresponds to HR$^+$ cells
(yellow in Fig. 2D), includes inhibition crosstalk as the second best
model, whereas in all other clusters this mechanism appears in fourth
place.  This finding suggests the possibility that the cell lines in
cluster~4 share a feature which is relevant to the ligands that also
appear in this cluster. This type of insight is made possible because
of the constraint we have imposed on the clusters.

\section{Discussion}

We have introduced a novel framework to cluster multi-indexed data
based on tensors that allows structural constraints to be incorporated
using algebraic relationships. This method can be used to extract
clusters directly from the data, and if an initial clustering which
may not satisfy the constraints is provided, it can find the closest
optimal partition that satisfies the constraints. A key advantage of
this framework is that it allows more control over the composition of
clusters than in many unsupervised methods, and allows the
clustering to be tailored to the requirements of the problem. The main
  limitation of this method is that it requires {\it complete} data
  (i.e., a measurement for every cell line/time/dose/ligand/molecule
  combination), which can be difficult to obtain. The metric used to compare data points could be adapted to deal with a small number of missing entries, but the method is unlikely to perform well for sparse data.

We applied this method on a dataset charting the response of
genetically diverse breast cancer cell lines to ligands.  We
identified both similarities (e.g., HER2$^{\rm{amp}}$) and
heterogeneities (e.g., TNBC) within clinical subtypes.  The
heterogeneity of our clustering analysis
(Fig.~\ref{fig:three-clusters}B) seems to be related to both the
mutational status of the cells as well as their response to
inhibitors. This result means that similar analyses in patient tissues
might be able to identify patients that respond differently to
therapeutic methods commonly used within a clinical subtype. By
analyzing clusters from all subtypes, we also showed that we cannot
attribute the dynamics of each data cluster with only one signalling
mechanism, which helps explain network model differences across cell
type.

The applicability of our method goes beyond the biological problem
presented here. It can be used in any context in which the
constraints on the clusters can be expressed in algebraic form (as
equalities and inequalities), such as when there are size restrictions
on the clusters, or to impose/prohibit particular combinations of data
beyond {\it must-link} and {\it cannot-link} constraints. For example,
this method could be used to construct optimal portfolios that comply
with rules about their composition~\cite{McNeil2015}, to help the
formation of teams that maximize members' preferences and are
compliant with skill requirements~\cite{Davis1971}, and to find
communities in networks with quotas, among others.  The presented
pipeline (a sophisticated and interpretable data analysis method that
feeds into a nonlinear modeling framework) will be ever more necessary
as increasingly more large-scale, comprehensive datasets become
available.

\section*{Author Contributions}

\vspace{-1ex}
AS, MBD and HAH developed the methodology, performed the analysis and wrote the manuscript. AS, MBD, BS and HAH designed the study. BS and MN provided datasets for analysis and interpretation.

\section*{Data Accessibility}

\vspace{-1ex}
The data set used in this analysis can be found in Ref.~\cite{Niepel2014}, we also give a link to the data in the SI. The Matlab code for the constrained clustering method can be found in the SI Section B.5 `Code for constrained clustering'. A link to the code for the mechanistic analysis can be found in the SI Section D.3 `Parameter estimation'. 

\section*{Funding Statement}

\vspace{-1ex}
AS and HAH acknowledge funding from the Royal Society International
Exchanges Scheme 2014/R1 IE140219. HAH gratefully acknowledges funding
from EPSRC Fellowship EP/K041096/1 and a Royal Society University
Research Fellowship. MBD acknowledges support from the Oxford-Emirates Data
Science Lab and James S. McDonnell Foundation Postdoctoral Program in
Complexity~Science/Complex~Systems Fellowship Award (\#220020349-CS/PD
Fellow).  

%\section*{Competing interests}
%
%The authors declare no competing interests.

\section*{Acknowledgements}

\vspace{-1ex}
We thank Mauricio Barahona, Sam Howison, Peter Mucha, Bernd
Sturmfels, Puck Rombach and Yan Zhang for discussions and comments.

\bibliographystyle{pnas}

%\end{document}

%
\newpage
\clearpage

% \onecolumn
\onecolumngrid
\appendix

\section*{Supplementary Information: 
Tensor clustering with algebraic
  constraints gives interpretable groups of crosstalk mechanisms in
  breast cancer}
  
  \setcounter{figure}{0}
\makeatletter 
\renewcommand{\thefigure}{S\@arabic\c@figure}
\makeatother

\section{Data}

\begin{figure}[bp]
  \centerline{\includegraphics[width=.8\textwidth]{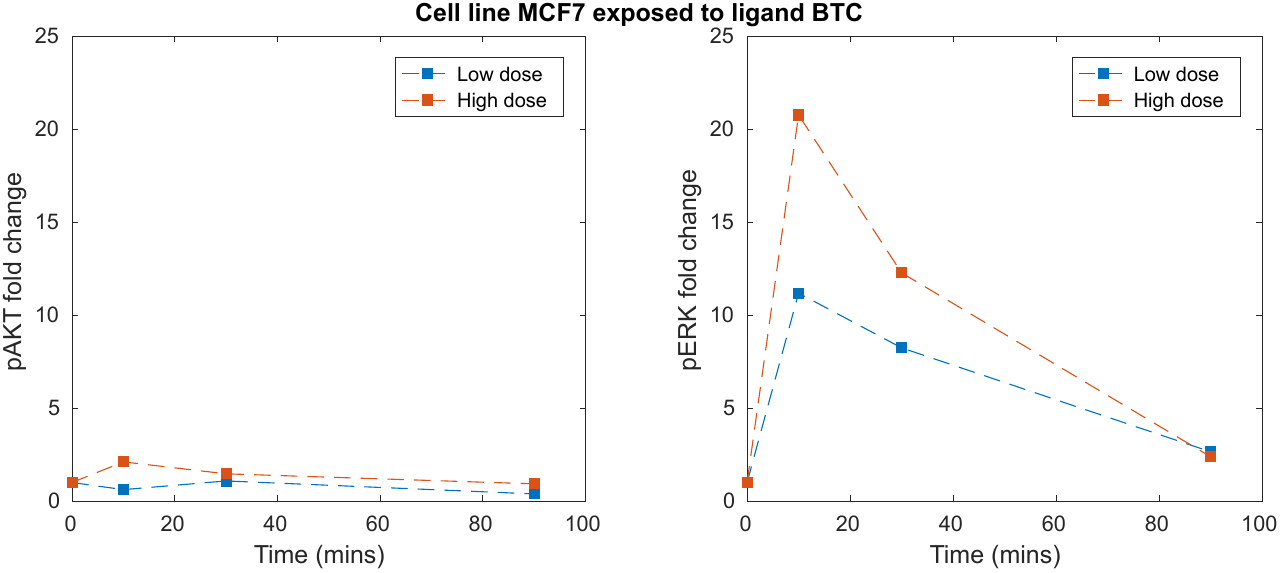}}
  \caption{Time course measurements of cell line MCF7 exposed to two
    doses of betacellulin (BTC).}
  \label{fig:example_timecourse}
\end{figure}

We analyse a data set first presented in Ref.~\cite{Niepel2014}, which is available for download at the webpage {http://lincs.hms.harvard.edu/niepel-bmcbiol-2014/}.  The
data consists of time course measurements (at 0, 10, 30, and 90
minutes) of the fold change in the phosphorylation levels of the
mitogen activated protein kinase (MAPK) pERK and the phosphoinositide
3-kinase (PI3K) pAKT in 36 breast cancer cell lines
(Table~\ref{tab:cell-lines}), each exposed to two doses (low:
$1\,$ng/ml and high: $100\,$ng/ml) of 14 different ligands
(Table~\ref{tab:ligands}).  Figure~\ref{fig:example_timecourse} shows
an example of the temporal response of pAKT and pERK in the cell line
MCF7 to doses of betacellulin.

The data set is complete in the sense that it contains measurements of
pAKT and pERK in every cell line, exposed to 2 doses of each of the 14
ligands at 4 time points. As a result, we are able to represent the
data as a $\times 36\times 14 \times 2 \times 3 \times 2$ tensor $\mbf{Z}$,
with entries $z_{ijptd}$. The index $p\in \{\text{pAKT},\,
\text{pERK}\}$ denotes which protein was measured, $i$ corresponds to
the cell lines in Table~\ref{tab:cell-lines}, $j$ corresponds to the
ligands in Table~\ref{tab:ligands}, $d \in \{1,\,100\}$ denotes the
doses, and $t\in\{10,\,30,\,90\}$ is time after the stimulus.

\begin{table}[tp]
  \begin{center}
    {\small
      \begin{tabular}{|l|l|}
        \hline
        {\bf Cell line} & {\bf Subtype} \\
        \hline
        MCF7 & HR+ \\
        SK-BR-3 & HER2amp \\
        MDA-MB-231 & TNBC \\
        AU-565 & HER2amp \\
        BT-20 & TNBC \\
        BT-474 & HER2amp \\
        BT-483 & HR+ \\
        BT-549 & TNBC \\
        CAMA-1 & HR+ \\            
        HCC-1395 & TNBC \\
        HCC-1419 & HER2amp \\
        HCC-1428 & HR+ \\
        HCC-1569 & HER2amp \\
        HCC-1806 & TNBC \\
        HCC-1937 & TNBC \\
        HCC-1954 & HER2amp \\
        HCC-202 & HER2amp \\
        HCC-38 & TNBC \\
        HCC-70 & TNBC \\
        Hs 578T & TNBC \\
        MDA-MB-134VI & HR+ \\
        MDA-MB-157 & TNBC \\
        MDA-MB-175VII & HR+ \\
        MDA-MB-361 & HER2amp \\
        MDA-MB-415 & HR+ \\
        MDA-MB-436 & TNBC \\
        MDA-MB-453 & TNBC \\
        MDA-MB-468 & TNBC \\
        T47D & HR+ \\
        UACC-812 & HER2amp \\
        UACC-893 & HER2amp \\
        ZR-75-1 & HR+ \\
        ZR-75-30 & HER2amp \\
        184-B5 & TNBC \\
        HCC-1187 & TNBC \\   
        HCC-1500 & HR+ \\
        %CF-10A & TNBC \\
        %CF 10F & TNBC \\
        %CF-12A & TNBC \\
        %U-4475 & TNBC \\
        %CC-2218 & HER2amp \\
        %CC-2157 & TNBC \\
        %CC-1599 & TNBC \\
        \hline
      \end{tabular}
    }
  \end{center}
  \caption{Breast cancer cell lines used in the data set~\cite{Niepel2014}.}
  \label{tab:cell-lines}
\end{table}

\begin{table}[tp]
  \begin{center}
    {\small
      \begin{tabular}{|l|l|}
        \hline
        {\bf Ligand name} & {\bf Abbreviation} \\
        \hline
        Betacellulin & BTC\\
        Epidermal Growth Factor & EGF\\
        Epiregulin & EPR\\
        Fibroblast Growth Factor (acidic) & FGF-1\\
        Fibroblast Growth Factor (basic) & FGF-2\\
        Hepatocyte Growth Factor & HGF\\
        Heregulin $\beta$1 & HRG\\
        Insulin-like Growth Factor 1 & IGF-1\\
        Insulin-like Growth Factor 2 & IGF-2\\
        Insulin & INS\\
        Nerve Growth Factor & NGF-beta \\
        Platelet Derived Growth Factor BB & PDGF-BB\\
        Stem Cell Factor & SCF \\
        Vascular endothelial growth factor A & VEGF165\\
        \hline
      \end{tabular}
    }
  \end{center}
  \caption{Ligands used in the data set~\cite{Niepel2014}.}
  \label{tab:ligands}
\end{table}

\section{Clustering}

We first describe how we normalize the experimental data for
clustering. Then we provide details for how we clustered directly from
the data, and then how we obtained and incorporated an initial
clustering.

\subsection{Normalization}

We normalize the data as follows. We scale all the pAKT and pERK
foldchange responses such that their average takes each the value
1. This normalization balances the effects of AKT and ERK, so that the
behavioural features and not the scale are the dominating features,
and to ensure we treat them with equal significance in our study. The
mean value (pre-normalization) across the AKT responses is 1.7754 and
that across the ERK responses is 11.4190.

\subsection{No prior clustering}

\begin{figure}[tp]
\centering
   \includegraphics[width=.5\textwidth]{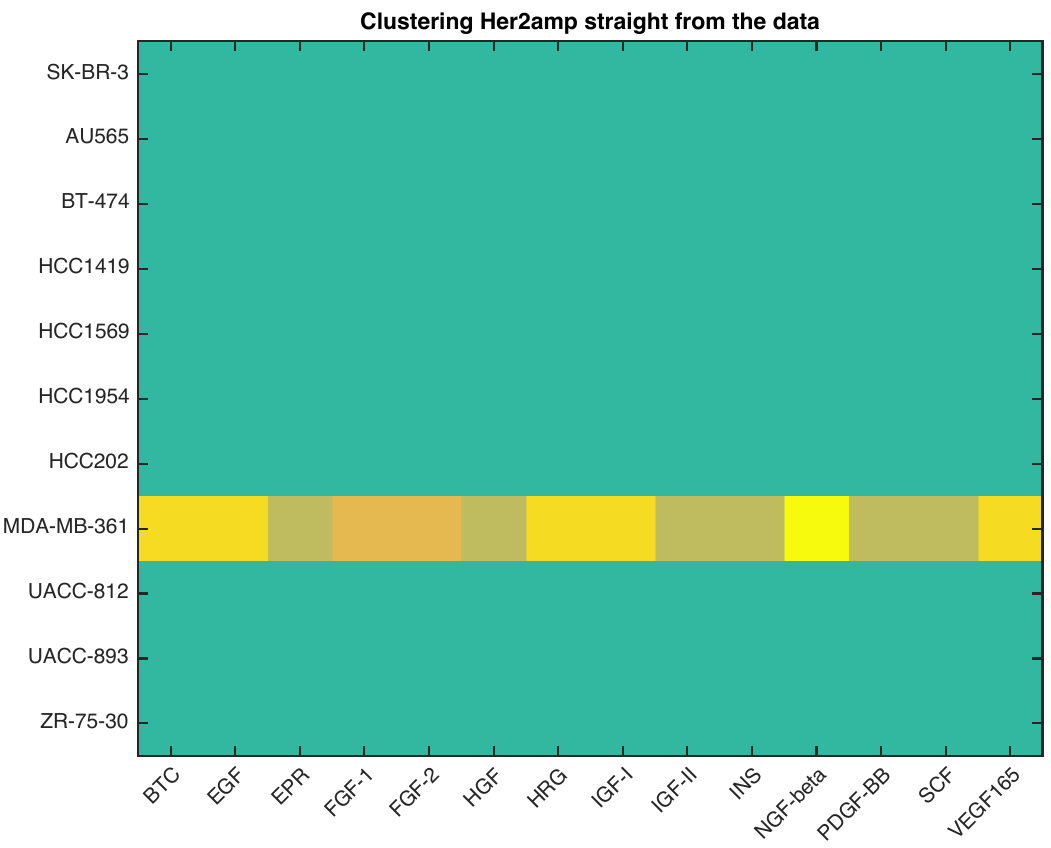}
  \caption{Clusters from the data of HER2$^{\rm amp}$ cell line.}
  \label{fig:2}
\end{figure}

As a summary statistic of the three clinical subtypes, we compute the
average distance score within each subtype. The score for the 11
HER2$^{\rm amp}$ cell lines is obtained as follows. There are $154 =
11 \times 14$ experiments in the dataset that involve HER2$^{\rm amp}$
cell lines, each consisting of 12 measurements. For each pair of
experiments, we find the dissimilarity between the 12 measurements,
using cosine dissimilarity
$$
1 - \frac{ \langle v_1, v_2 \rangle }{\| v_1 \|_2 \| v_2 \|_2 } ,
$$
where $\langle \cdot, \cdot \rangle$ is the usual inner product in a real vector space and $\| \cdot \|_2$ is the Euclidean norm. 
The average
dissimilarity is obtained by averaging these pairwise distances across
the ${{ 154 \choose 2 }} = 11781$ pairs. Similarly for the HR$^+$ cell
lines and for the TNBC cell lines. The averages obtained are 0.086 for
HER2$^{\rm amp}$, 0.334 for HR$^+$ and 0.224 for TNBC.  The
partitioning of the HER2$^{\rm amp}$ cell lines is given in
Fig.~\ref{fig:2}.

\subsection{Pre-existing clusters}

\subsubsection{Computing pre-existing clusters}

 We find an initial clustering of the experiments. For this initial
 clustering, we label each experiment by a single index. The data for
 the $i$th experiment is:
\begin{equation}
        \mbf{\widetilde{Z}}(i,:) =
        \left[\left.\mbf{AKT}^{1}_i\right|\left. \mbf{ERK}^{1}_i\right|
        \left.\mbf{AKT}^{100}_i\right|\mbf{ERK}^{100}_i\right],
        \label{eq:descriptor}
\end{equation}
where $\mbf{AKT}^{1}_i$ is the normalised time series of fold-change
response of pAKT under dose $1$ng/ml, and so on. We compute the
$504\times504$ similarity matrix $\mbf{\widetilde{S}}$, in which
$s_{ij}$ indicates the cosine similarity of experiments $i$ and $j$:
\begin{equation}
 s_{ij} = \frac{\inprod{\tilde{\mbf {z}}_i}{\tilde{\mbf
       {z}}_j}}{\norm{\tilde{\mbf {z}}_i}_2\norm{\tilde{\mbf
       {z}}_j}_2} = \cos{\left(\tilde{\mbf {z}}_i,\tilde{\mbf {z}}_j
   \right)}.
\label{eq:similarity}
\end{equation}
where $\tilde{\mbf {z}}_i=\mbf{\widetilde{Z}}(i,:)$ and $\tilde{\mbf
  {z}}_j=\mbf{\widetilde{Z}}(j,:)$.  The entries of $\tilde{\mbf
  {z}}_i$ and $\tilde{\mbf {z}}_j$ are nonnegative, which means that
$s_{ij} \in [0,1].$ If $s_{ij}=1$, experiments $i$ and $j$ have an
{\it identical} response to the treatments in both AKT and ERK (up to
a scaling constant). When $s_{ij}=0$, the data for the experiments are
orthogonal. The task of clustering the experiments faces two
challenges: the number of clusters is not known a priori, and the
matrix $\mbf{\widetilde{S}}$ is full matrix and is noisy due to
experimental error.  To tackle these challenges we use a combination
of tools from manifold learning and network science.  We create a
network (graph) in which each of the 504 experiments is represented by
a node, and where connections exist between similar experiments. We
first define the dissimilarity matrix $\mbf{D}$ ($d_{ij} = 1 -
s_{ij}$).  We then use the {\it Relaxed Minimum Spanning Tree} (RMST)
algorithm~\cite{Vangelov2014,Beguerisse2013,Beguerisse2014}, which
extracts a network representation from high-dimensional point clouds
(in this case the $\tilde{\mbf {z}}_i$) that are embedded in a lower
dimensional manifold.

Specifically, the algorithm creates an undirected, unweighted network
with an edge between $i$ and $j$ if they are neighbors in a minimum
spanning tree (MST) from $\mbf{D}$. The algorithm adds extra edges to
the network if they are consistent with the continuity of the data,
i.e. if the distances between the points in $\mbf{D}$ is comparable to
their separation in the MST and is consistent with the continuity of
the data, according to the equation $$d_{ij} < \rm{mlink}_{ij} +
\frac{1}{2}(k_i + k_j).$$ Here $\rm{mlink}_{ij}$ is the maximal edge
weight in the MST path connecting $i$ to $j$, and $k_{i}$ is the the
distance to the nearest neighbour of $i$ (i.e., the minimum value on
the $i$th row of $\mbf{D}$, excluding $d_{ii}$). Basically, what the
RMST algorithm does is allows edges to be added to the MST (it
`relaxes' the MST), so that we obtain an network description of high
dimensional data that is embedded on a lower dimensional manifold.

\begin{figure}[tp]
\centering
   \includegraphics[width=.9\textwidth]{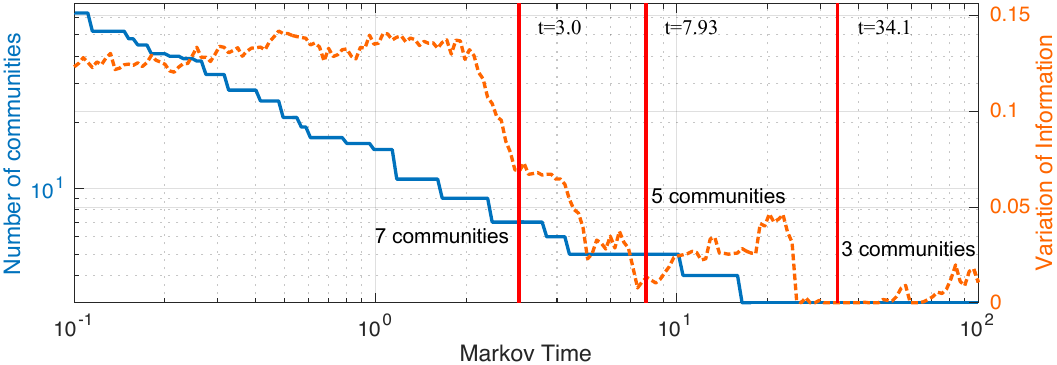}
  \caption{Number of communities and variation of information for the
    network obtained from RMST similarity graph. } \label{fig:sim}
\end{figure}

Once we have obtained the network from the similarity matrix, we
extract communities using the Markov Stability (MS) community
detection algorithm~\cite{Delvenne2010,Delvenne2013}. This method
employs continuous time random walks of varying duration (Markov time)
to extract communities of the network at different levels of
resolution. Shorter Markov times produce small communities, whereas
longer Markov times lead to coarse partitions of the network.
Obtaining the optimal partition of a network into communities is an NP
complete problem, so MS uses heuristics to find communities. Because
there is no guarantee of finding a global optimum, MS repeats the
heuristic search 100 times for each Markov time. The variability in
each set of 100 solutions is measured with the Variation of
Information (VI)~\cite{Meila2007}. A low value of the VI for a Markov
time indicates that the solutions obtained are similar to each other,
we take this similarity as a sign that there is a robust partition of
the network for this Markov time.  In Fig.~\ref{fig:sim} we show that
the network $A$ has a robust partition into 3, 5 and 7 communities.

\subsubsection{Structured clusters from pre-existing clusters}

\begin{figure}[tp]
\centering
  \includegraphics[width=.5\textwidth]{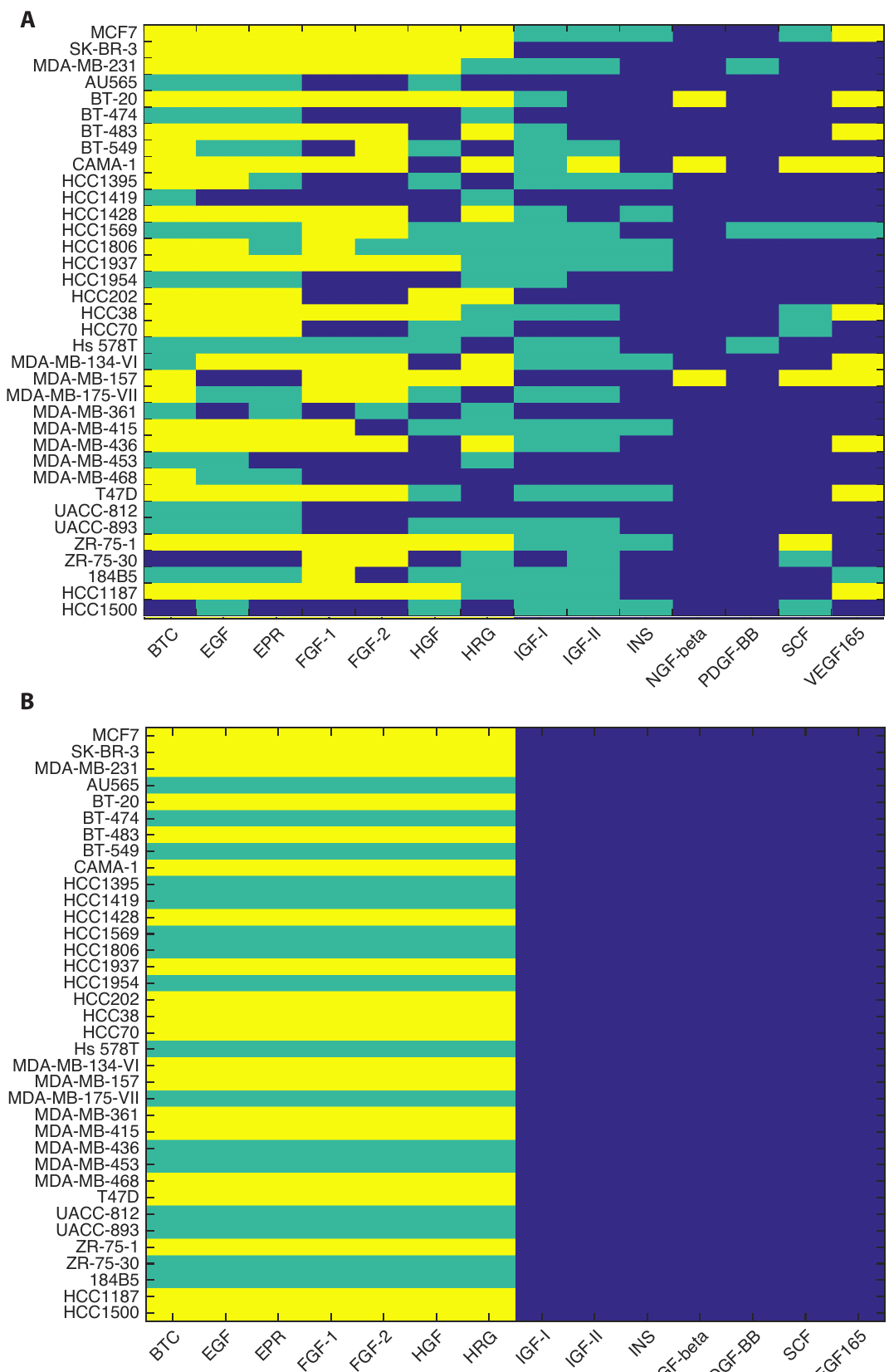}
  \caption{A. The clustering assignments are represented by yellow,
    green and blue squares. B. The tensor clustering are a close
    approximation to the clusters obtained with MS.}
  \label{fig:1}
\end{figure}

We present the clustering assignments after using MS to obtain our
initial partition into clusters. The pre-existing and structured
cluster results for three clusters (Fig.~\ref{fig:1}), five clusters
(Figs.~3D, \ref{fig:1}) and seven clusters (Fig.~\ref{fig:seven}).

\begin{figure}[tp]
\centering
   \includegraphics[width = .7\textwidth]{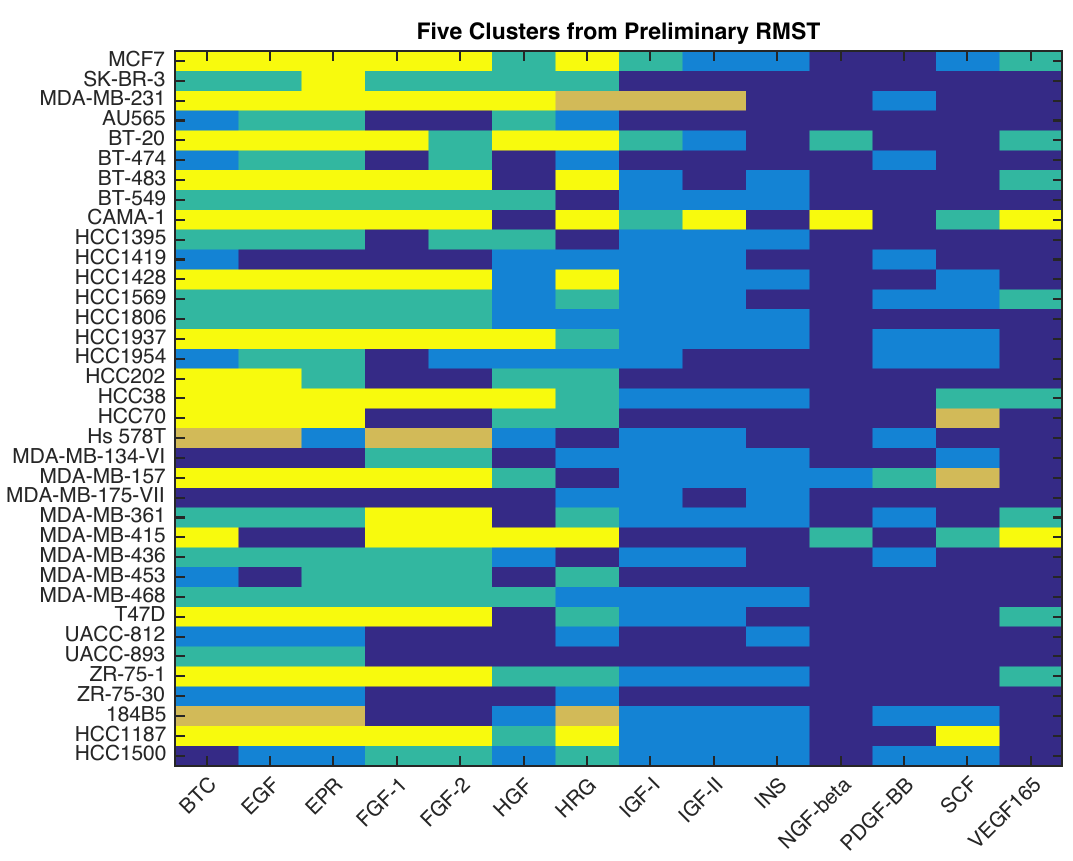}
  \caption{Five clusters from Markov Stability } \label{fig:five}
\end{figure}

A finer clustering into seven groups (see Fig.~\ref{fig:seven})
divides the ligands into three groups: high response (yellow, green,
brown) \{BTC, EGF, EPR, FGF-1, FGF-2, HGF, HRG\} and lower response
(blues) \{IGF1, IGF2, INS\} and \{ NGF-$\beta$, PDGF-BB, SCF,
VEGF175\}. The assignment of cell types is remarkably similar to the
five cluster results. The exception cell lines are: HCC1419 (green to
brown), and ZR-75-30 (green to the new seventh cluster).  Given this
consistency, we restrict to mechanistic interpretation of the five
cluster case.

\begin{figure}[tp]
  \centerline{ \includegraphics[width=.45\textwidth]{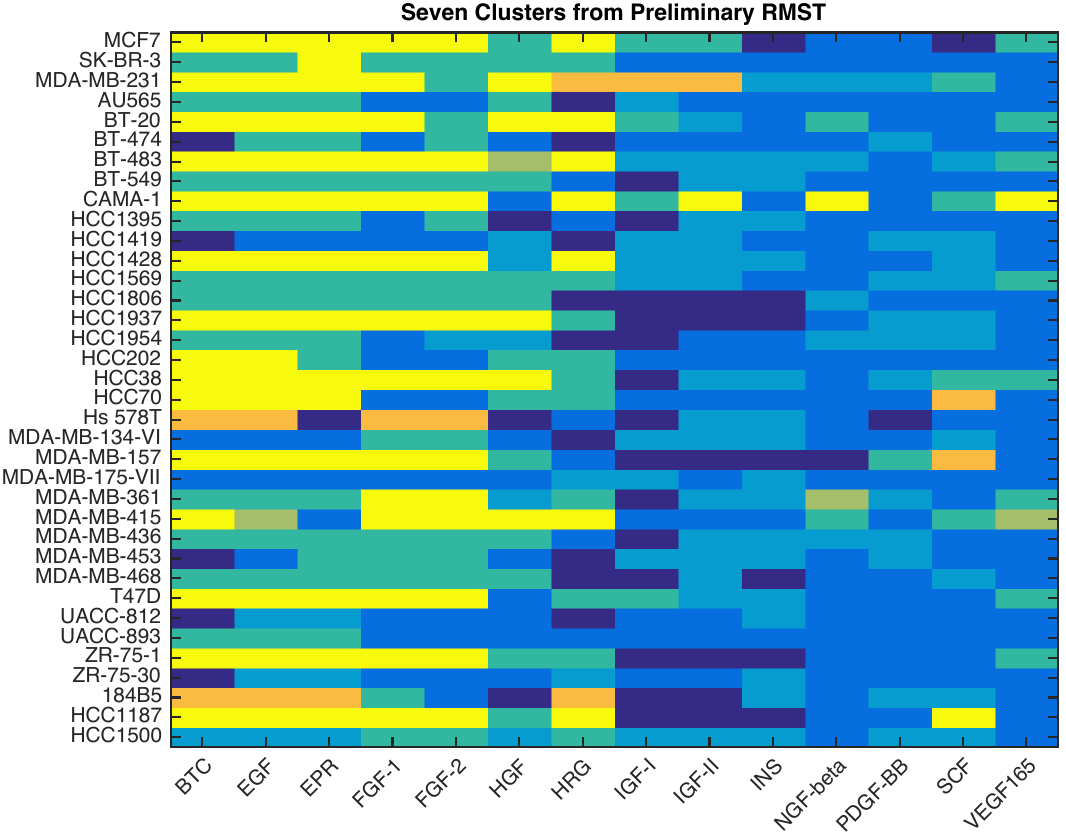}
    \includegraphics[width =.45\textwidth]{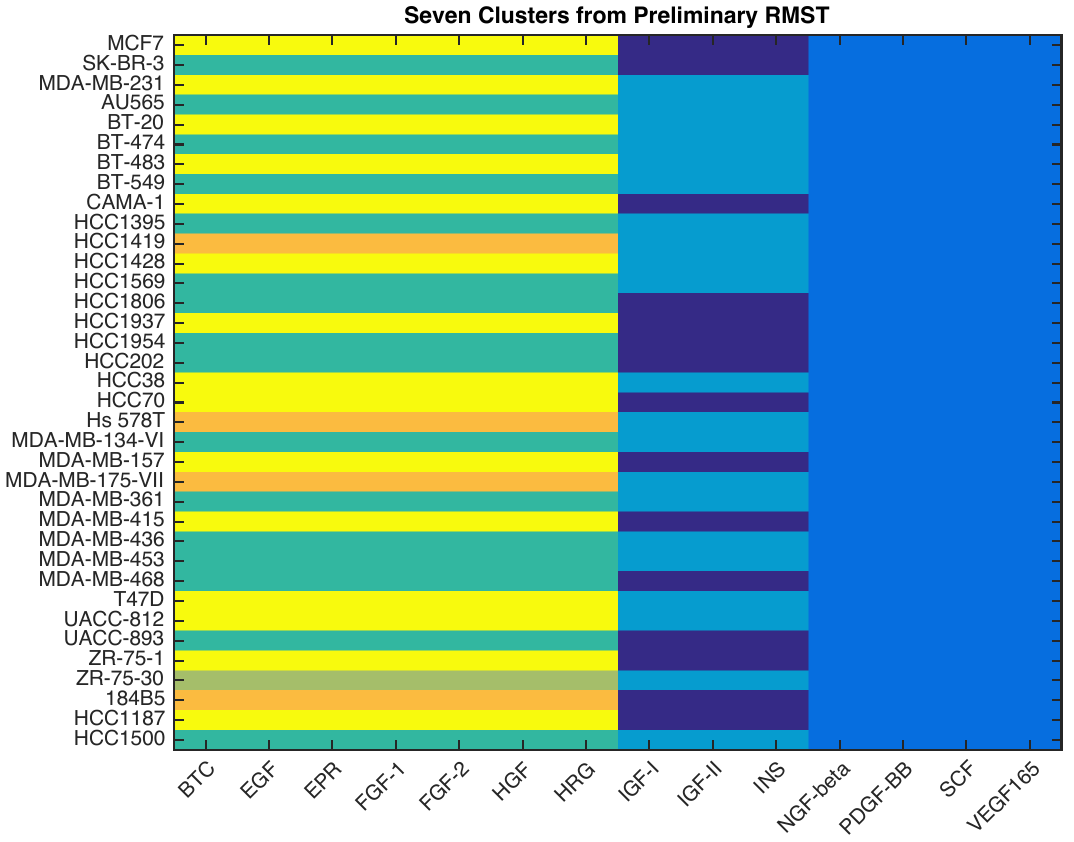}}
  \caption{Seven clusters from Markov Stability.  }\label{fig:seven}
\end{figure}

\subsection{Comparison of implementations}

\begin{figure}[tp]
\centering
   \includegraphics[width = .6\textwidth]{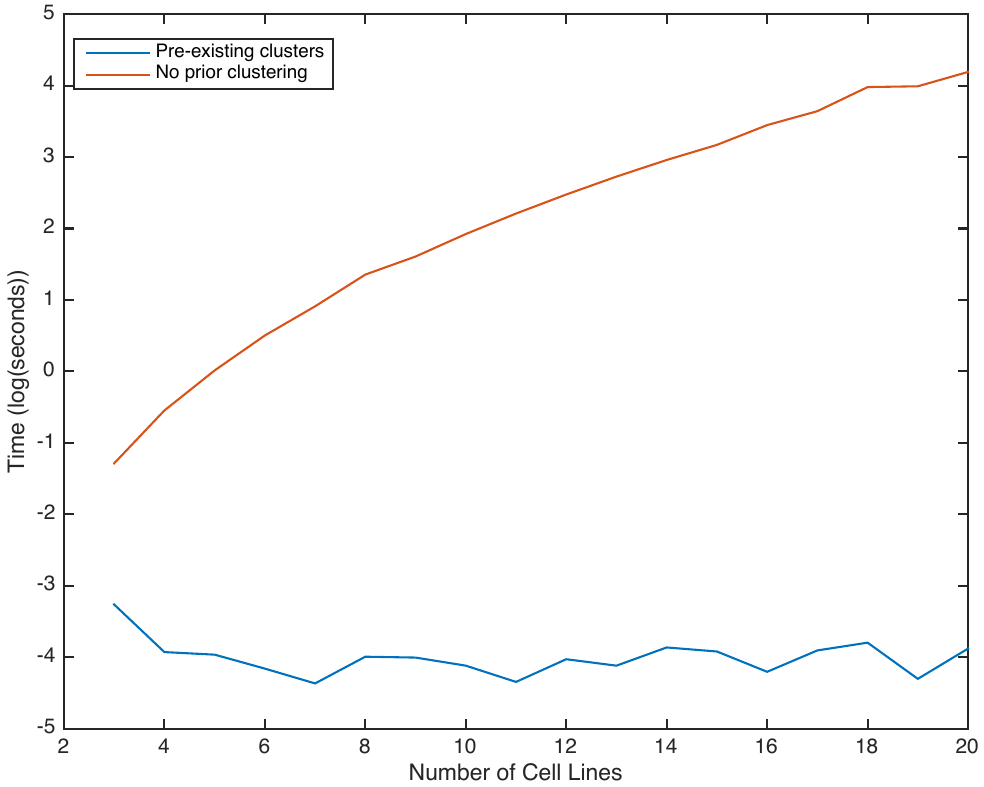}
  \caption{Computational complexity of two implementations with eight
    ligands and a varying number of cell lines.  } \label{fig:cplex}
\end{figure}

When we employ our method assuming no initial clusterings, the integer
optimization is learning the values, 0 or 1, for an array of size $c
\times c \times l \times l$ (where $c = \#$cell lines and $l = \#$
ligands). When we use an initial clustering, CPLEX uses the same
branch and cut algorithm on an array of size $c \times l \times k$,
where $c=\#$cell lines, $l=\#$ligands and $k=\#$clusters, so the array
is much smaller for the pre-existing clustering implementation. We
show a comparison of the performance of the two implementations in
Fig.~\ref{fig:cplex}

\subsection{Code for constrained clustering}

The following pieces of Matlab code generate the constraints on the tensors that encode the clustering assignment, under the rectangle clusterings condition.  Assume we have a dataset of size $n \times m \times \cdots \times o$ and we wish to cluster into rectangular-shaped clusters using the first two indices. First, reshape the data into a three-dimensional array, \verb|Tensr|, by vectorizing the 3rd, 4th, \ldots, indices into a multi-index of size $p$. Then, the following code makes the similarity tensor with respect to cosine dissimilarity.  

\begin{verbatim}
[n m p] = size(Tensr);
C = zeros(n,n,m,m);
for i = 1:n; for j = 1:n; for k = 1:m; for l = 1:m;
                v1 = zeros(p,1); v2 = zeros(p,1);
                for ii = 1:p;
                    v1(ii) = Tensr(i,k,ii); v2(ii) = Tensr(j,l,ii);
                end
                C(i,j,k,l) = 1 - dot(v1,v2)/(norm(v1,2)*norm(v2,2));
end end end end
\end{verbatim}

Next we encode the constraints on the tensor which encodes the partition of the data set. It is constructed as a sparse array, i.e. by specifying row and column indices \verb|Vrow| and \verb|Vcol|, and values \verb|Vval| of all non-zero entries. The lower and upper bounds for each linear constraints are organized into the vectors \verb|lb| and \verb|ub| respectively.
\begin{verbatim}
t = 1;
for i = 1:n; for j = 1:n; for k = 1:m; for l = 1:m;
                if (i ~= j) || (k ~= l);
                    W = vectorize(n,i,j,k,l);
                    Vrow(w) = t; Vcol(w) = W;Vval(w) = 1;
                    w = w+1;
                    W = vectorize(n,j,i,l,k);
                    Vrow(w) = t; Vcol(w) = W; Vval(w) = -1;
                    w = w+1;
                    lb(t) = 0; ub(t) = 0;
                    t = t+1;
                end
                if (k ~= l);
                    W = vectorize(n,i,j,k,l);
                    Vrow(w) = t; Vcol(w) = W; Vval(w) = 1;
                    w = w+1;
                    W = vectorize(n,i,j,l,k);
                    Vrow(w) = t; Vcol(w) = W; Vval(w) = -1;
                    w = w+1;
                    lb(t) = 0; ub(t) = 0;
                    t = t+1;
                % opposite diagonals must be same
                end
                if (i ~= j);
                    W = vectorize(n,i,j,k,l);
                    Vrow(w) = t; Vcol(w) = W; Vval(w) = 1;
                    w = w+1;
                    W = vectorize(n,j,i,k,l);
                    Vrow(w) = t; Vcol(w) = W; Vval(w) = -1;
                    w = w+1;
                    lb(t) = 0; ub(t) = 0;
                    t = t+1;
                end
                if (k ~= l); 
                    W = vectorize(n,i,j,k,l);
                    Vrow(w) = t; Vcol(w) = W; Vval(w) = 1;
                    w = w+1;
                    W = vectorize(n,j,i,k,k);
                    Vrow(w) = t; Vcol(w) = W; Vval(w) = -1;
                    w = w+1;
                    lb(t) = 0; ub(t) = 1;
                    t = t+1;
                % vertical conditions
                end
                if (i ~= j);
                    W = vectorize(n,i,j,k,l);
                    Vrow(w) = t; Vcol(w) = W; Vval(w) = 1;
                    w = w+1;
                    W = vectorize(n,i,i,k,l);
                    Vrow(w) = t; Vcol(w) = W; Vval(w) = -1;
                    w = w+1;
                    lb(t) = 0; ub(t) = 1;
                    t = t+1;
                % horizontal conditions
                end
end end end end

for i = 1:n; for k = 1:m;
        W = vectorize(n,i,i,k,k);
        Vrow(w) = t; Vcol(w) = W; Vval(w) = 1;
        w = w+1;
        lb(t) = 0; ub(t) = 0;
        t = t+1;
end end

for i1 = 1:n; for i2 = 1:n; for i3 = 1:n;
for k1 = 1:m; for k2 = 1:m; for k3 = 1:m;
if (((i1 ~= i2) || (k1 ~= k2)) && ((i2 ~= i3) || (k2 ~= k3)) && ((i1 ~= i3) || (k1 ~= k3)));
             W = vectorize(n,i1,i3,k1,k3);
             Vrow(w) = t; Vcol(w) = W; Vval(w) = -1;
             w = w+1;       
             W = vectorize(n,i1,i2,k1,k2);
             Vrow(w) = t; Vcol(w) = W; Vval(w) = 1;
             w = w+1;  
             W = vectorize(n,i2,i3,k2,k3);
             Vrow(w) = t; Vcol(w) = W; Vval(w) = 1;
             w = w+1;
             lb(t) = 0; ub(t) = 2;
             t = t+1;
end end end end end end end
maxt = t-1;
V = sparse(Vrow,Vcol,Vval,maxt,n*n*m*m);
\end{verbatim}

Now complete the optimization using CPLEX, feeding the above constraints into the model alongside a penalty parameter that controls the number of clusters. The function \verb|vectorize|, used above, combines the multi-index into a single index:
\begin{verbatim}
function t = vectorize(n,i,j,k,l);
m = 14; % for example
i1 = n*m*m*(i-1); j1 = m*m*(j-1); k1 = m*(k-1); l1 = l;
t = i1+j1+k1+l1;
\end{verbatim}

For clustering based on pre-existing clusters, we generate the constraints imposed using the following Matlab code. Assume we are clustering based on 2 indices of sizes $n$ and $m$, and that the preliminary clustering returned \verb|nclu| clusters. As before, the lower and upper bounds on the conditions are organized into the vectors \verb|lb| and \verb|ub|.
 
\begin{verbatim}
w = 1; t = 1;

for i = 1:n; for j = 1:m;
	for k = 1:nclu;
            W = nclu*m*(i-1) + nclu*(j-1) + k;
            % turn i,j,k coordinates into a vector
            Vrow(w) = t; Vcol(w) = W; Vval(w) = 1;
            w = w+1;
        end
        lb(t) = 1; ub(t) = 1;
        t = t+1;
end end
 the vector at position (i,j) has entries which sum to 1

for i = 1:n; for j = 1:n; for k = 1:m; for l = 1:m;
                if (i ~= j) && (k ~= l) ;
                for r = 1:nclu;
                    W = nclu*m*(i-1) + nclu*(k-1) + r;
                    Vrow(w) = t; Vcol(w) = W; Vval(w) = 1;
                    w = w+1; 
                    W = nclu*m*(j-1) + nclu*(l-1) + r;
                    Vrow(w) = t; Vcol(w) = W; Vval(w) = 1;
                    w = w+1; 
                    W = nclu*m*(i-1) + nclu*(l-1) + r;
                    Vrow(w) = t; Vcol(w) = W; Vval(w) = -1;
                    w = w+1;
                    lb(t) = -1; ub(t) = 1; 
                    t = t+1; 
                end end
end end end end
maxt = t-1;
 make sparse array of the constraints:
Vnew = sparse(Vrow,Vcol,Vval,maxt,n*m*nclu);
\end{verbatim}

\section{Models}

\subsection{Models of the MAPK pathway}

The MAP Kinase pathway has been widely studied. Models have focused on
various features of the cascade, such as its three-tier
phosphorylation feedback structure~\cite{Heinrich2002,Huang1996}. The
activation profile of these kinases is directly related to cellular
decisions and
fates~\cite{kriegsheim:2009,Marshall1995,Purvis:2013dd,Chen239}. The
AKT pathway has been modeled by EGF-dependent activation, including
phosphorylation of AKT (pAKT) and its downstream intracellular
proteins~\cite{Liepe2013}. Few models of crosstalk between ERK and AKT
exist. One model was created for studying PC12 cells, and they found
that AKT acts as a low-pass filter which decouples the EGF
signal~\cite{Fujita2010}. Another model was created to study HEK293
cells in the presence of a MEK inhibitor; they found crosstalk is
reinforced between Ras and PI3K~\cite{Won01062012}. Another model
found that JNK is regulated by AKT and MAPK feedbacks in these
pathway~\cite{Fey:2012ge}. Chen and co-authors constructed and
analyzed an ErbB model focusing on the receptor dynamics and early
activation response of the MAPK and AKT pathways in response to
ligands in A431 or H1666 cells \cite{Chen239}; however their model was
unidentifiable, meaning there were an infinite number of parameter
values to fit the data \cite{Chen239}. Weakly activated models of MAPK
activation cascades with optimal amplification under a variety of
stimuli were analyzed in~\cite{Beguerisse2016}.

\subsection{All wiring diagrams}

We consider all possible wiring diagrams to describe the interactions
between the receptor, the Erk pathway and the Akt pathway. These can
be written as a wiring-diagram where we assume an arrow exists between
the ligand $L$ and receptor $R$:
\begin{center}
  {\tiny
    \begin{tikzpicture}
      \node[rectangle,fill=blue!70!green!60!,rounded corners] (E) at (2,0) {pErk};
      \node[rectangle,fill=yellow!70!,rounded corners] (L) at (0,3) {Ligand};
      \node[rectangle,fill=blue!20!green!60!,rounded corners] (R) at (0,2) {Receptor};
      \node[rectangle,fill=red!60!,rounded corners] (A) at (-2,0) {pAkt};
      \draw[->,draw=black!50,line width=.75pt] (L) -- (R);
      \draw[-,draw=black!50,line width=.75pt] (R) -- (E);
      \draw[-,draw=black!50,line width=.75pt] (R) -- (A);
      \draw[-,draw=black!50,line width=.75pt] (A) -- (E);
      \draw[-,draw=black!50,line width=.75pt] (E) -- (A);
    \end{tikzpicture}
  }
\end{center}

We first consider all possible network topologies with interactions
between the variables $R$, $E$ and $A$. There are three possibilities
for the directed interaction from one variable to another: positive
($\rightarrow$), negative ($\dashv$), or no significant interaction
(no arrow). There are six potential directed interactions in the
network which give a total of $3^6 = 729$ networks.

Many of these networks can be ruled out. The data shows some response
in pERK and pAKT for each stimuli, therefore we require that both pERK
and pAKT have at least one arrow coming into each of these (to ensure
a response). Given this restriction from the data, we can eliminate
many networks, for example a network where the receptor inhibits pAKT
and pERK is biologically infeasible. We also do not have dynamic data
of the phosphorylated receptor, therefore we cannot distinguish a
network topology that has an arrow feeding back from the phopho-form
to the receptor; thus we fix the interaction from pERK to R and pAKT
to R to none (no arrow). All of these restrictions produce
Table~\ref{tab-notation}.

\begin{table}[tp]
  \begin{center}
    {\small
      \begin{tabular}{|c|c|c|}
        \hline  {\bf Number of arrows}  & {\bf Total number of networks} & {\bf Number of networks we consider} \\
        \hline
        All & $3^4 = 81$                      & 15  \\
        $0$ &  $\binom{4}{0}\cdot 2^0 = 1$    & None  \\
        $1$ &  $\binom{4}{1}\cdot 2^1 = 8$    & None  \\
        $2$ &  $\binom{4}{2}\cdot 2^2 = 24$   & 3  \\
        $3$ &  $\binom{4}{3}\cdot 2^3 = 32$   & 8  \\
        $4$ &  $\binom{4}{4}\cdot 2^4 = 16$   & 4  \\
        \hline
      \end{tabular}
    }
  \end{center}
  \caption{In total, we consider 15 network topologies: those that are
    biologically plausible given the data. There are different
    possible kinetics for each model: mass-action or
    Michaelis-Menten. This gives $15 \times 2 = 30$ possible
    models. Furthermore, seven of these include inhibition ($\dashv$)
    which we model via either: blocking or removal (described in the
    text). Accounting for all our kinetic models gives a total of 44
    models.}
  \label{tab-notation}
\end{table}

Based on these wiring diagrams, we now consider different possible
kinetics for each arrow, summarized in the Table, and described in the
next subsection.

\subsection{Construction of mechanistic models}

We construct systems of ordinary differential equation models to
describe interaction dynamics between the receptor (R), pAKT (A) and
pERK (E). The equation describing the evolution of the phosphorylated
receptor is $dR/dt = \alpha L (R_{\rm{tot}} - R) - \delta R $. The
total amount of receptor, $R_{\rm{tot}}$, is estimated from the
receptor abundance data. The unphosphorylated
receptor is given by $(R_{\rm{tot}} - R)$. The parameter $\alpha$
determines the rate at which it is phosphorylated by the ligand dose
($L = 1$ or $100$ ng/ml). The time evolution of pERK and pAKT can be
activated by $R$. There are two other equations, $dE/dt$ and $dA/dt$,
which describe the change in the phosphorylation (in fold change) of
$A$ and $E$ with respect to time. Crosstalk between the ERK and AKT
pathways is encoded by interactions between pERK and pAKT, which can
either activate or inhibit the other pathway. Activation terms are
modeled using either mass action or Michaelis-Menten kinetics. We
consider two types of inhibition: blocking through a saturation term
or through a removal term using mass action kinetics.

Without data from receptor dynamics, we write the change in
phosphorylated receptor (in arbitrary units) as a function of time as:
$$R' = \alpha L (R_{tot} - R) - \delta R,$$ where $\alpha L (R_{tot} -
R)$ describes the fraction of non-phosphorylated receptor that becomes
phosphorylated at some rate proportional to the ligand $L$, and
$\delta$ is the rate at which $R$ is de-phosphorylated.

The other two equations, $E'$ and $A'$ describe the change in the
phosphorylation (in fold change) of $A$ and $E$ with respect to
time. These equations change based on the assumed interactions, each
different set of equations describes a different mechanistic model.

We assume that activation is either via mass-action kinetics or
Michaelis-Menten kinetics, and that inhibition is either via removal
or blocking. For example, if we consider the model $R \rightarrow E
\rightarrow A, \quad R \dashv A$, we can write this as:
\begin{eqnarray}
  R' &=& \alpha L (R_{tot} - R) - \delta R,\\ E' &=& \frac{k_1
    R}{\textcolor{blue}{K_{m1} + R}} - \delta E,\\ A' &=& \frac{k_2
    E}{\textcolor{blue}{K_{m2} + E}} - {\textcolor{red}{k_3 AR}} -
  \delta A,
\end{eqnarray}
where the blue denotes the Michaelis-Menten term (ignoring the blue is
mass-action), and the red term $k_3 A E$ describes inhibition of $A$
as a removal interaction. However, when $A$ is inhibiting by blocking,
now the red term is written in a the following form:

\begin{eqnarray}
  R' &=& \alpha L (R_{tot} - R) - \delta R,\\ E' &=& \frac{k_1
    R}{\textcolor{blue}{K_{m1} + R}} - \delta E,\\ A' &=& \left(
  \frac{k_2 E}{\textcolor{blue}{K_{m2} + E}} \right)
  \left({\textcolor{red}{\frac{k_3}{k_3+R}}}\right) - \delta A.
\end{eqnarray}

\begin{figure}[tp]
  \centerline{
  \includegraphics[width=0.8\textwidth]{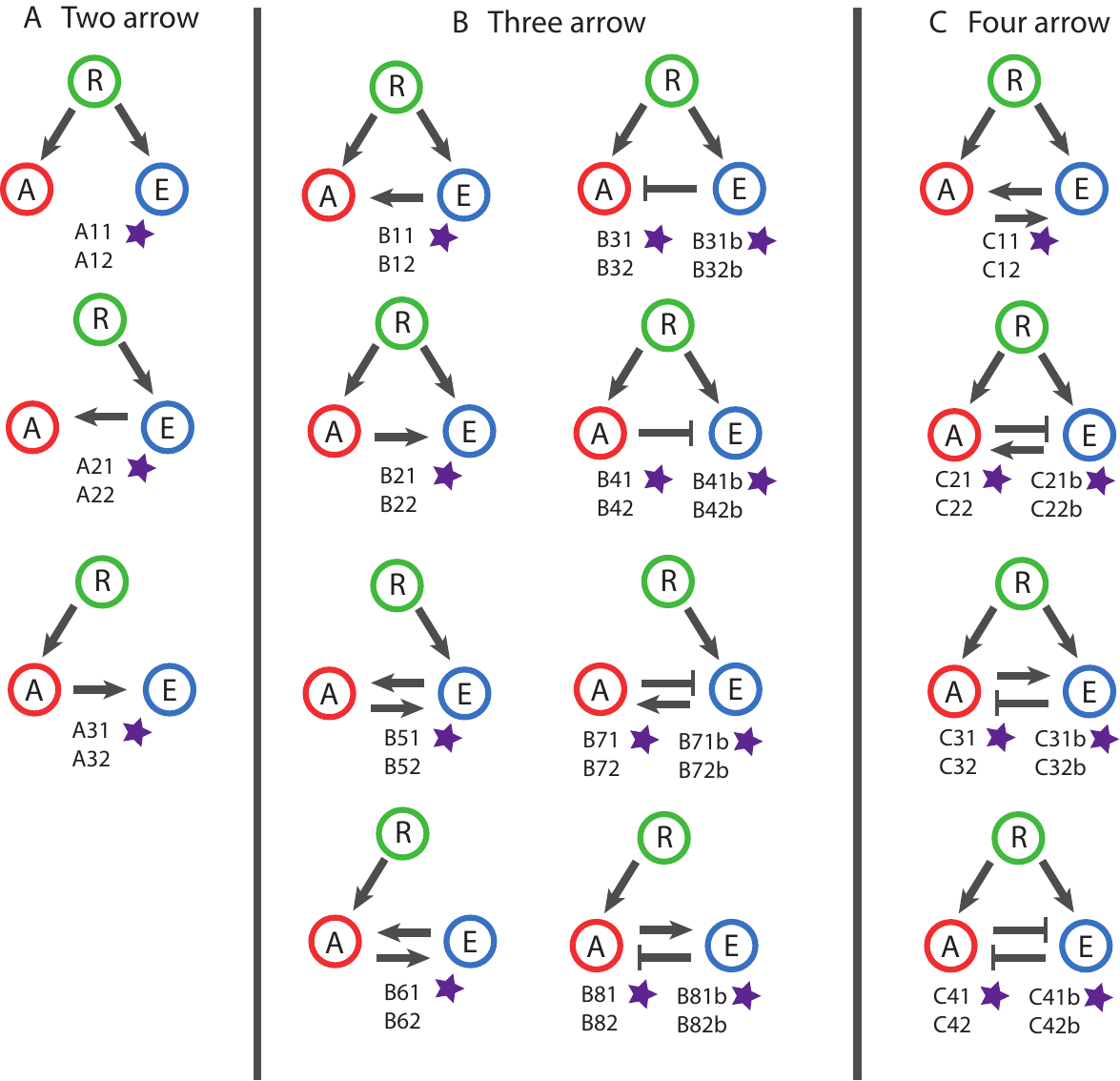}}
  \caption{ Mechanistic models of breast cancer cell lines. The name
    of network model is indexed first by whether it is a two arrow
    (A), three arrow (B), or four arrow (C) model. The second index
    assigns a number to each network topology. Each network can be
    further subdivided to describe a model with mass action kinetics
    (1 in third index) and a model with Michaelis-Menten kinetics (2
    in third index). Any network with inhibition ($\dashv$) has 4
    models considered: mass action removal (1 in third index),
    Michaelis-Menten removal (2 in third index), mass action blocking
    (1b in third index), or Michaelis-Menten removal (2b in third
    index). Stars next to the model name are globally structurally
    identifiable models, all other models are locally structurally
    identifiable. }
  \label{fig-cells}
\end{figure}

We summarize the 44 models analyzed in more detail in
Fig.~\ref{fig-cells}.

\section{Model identification}

\subsection{Estimating the total abundance of receptor, $R_{\rm{tot}}$} 

Our mechanistic analysis requires us to estimate the total abundance
of receptor (both phosphorylated and unphosphorylated) prior to the
addition of any ligand. We do this based on the Receptor Abundance Data from \cite[Figure 1]{Niepel2013}. Each ligand has one or more receptors associated to it, as shown in
Table~\ref{tab:recep}.

\begin{table}[tp]
\begin{center}
  \begin{tabular}{| l | l |}
    \hline
    Ligand & Associated Receptors \\ \hline \hline
    BTC & ErbB1, ErbB4 \\ \hline
    EGF & ErbB1 \\ \hline
    EPR & ErbB1, ErbB4 \\ \hline
    FGF-1 & FGFR-1, FGFR-2, FGFR-3, FGFR-4 \\ \hline
    FGF-2 & FGFR-1, FGFR-2, FGFR-3, FGFR-4 \\ \hline
    HGF & cMET \\ \hline
    HRG & ErbB4 \\ \hline
    IGF-1 & IGF1R \\ \hline
    IGF-2 & IGF1R, IGF2R \\ \hline
    INS & InsR \\ \hline
    NGF-beta & TrkA \\ \hline
    PDGF-BB & PDGFRa, PDGFRb \\ \hline
    SCF & c-Kit \\ \hline
    VEGF165 & VEGFR-1, VEGFR-2, VEGFR-3 \\ \hline
  \end{tabular}
\end{center}
\caption{Receptors associated to each ligand}
\label{tab:recep}
\end{table}

Each experiment involves a cell line and a ligand. For each cell
line/ligand pair, we estimate the receptor abundance by averaging the
receptor abundances for that cell line, for each of the receptors
associated to the ligand. We do this averaging over all receptor, cell
line pairs for which we have data.

We then estimate the receptor abundance for a cluster by averaging the
values obtained above for each cell line, ligand pair that is present
in the cluster.

\subsection{Identifiability analysis}
Before estimating model parameters from the data, we determine whether
a model is identifiable. Models that are globally identifiable have
parameters that are uniquely identifiable under ideal data
conditions. Models that are locally identifiability have a finite
number of indistinguishable parameter values. Since we only have
time-course measurements for $A$ and $E$, we use a differential
algebra approach for eliminating the species $R$. We test
identifiability using the algorithm {\tt DAISY} \cite{DAISY}. All of
our models are locally, if not globally identifiable given the
experimental data. Globally identifiable models are denoted by brown
boxes in Fig.~\ref{fig-cells}.

\subsection{Parameter estimation}
We estimate parameters using the average time-course for each
cluster. The model simulated at a parameter vector $\btheta$ gives a
vector of model predictions of $E$ and $A$ at dose $L$ and time point
$t_j$ and data $\mathcal{D}$ is the set of normalised measurements of
$\hat{E}$ and $\hat{A}$. The squared sum of errors of the model with
parameter set $\btheta$ is:
\begin{align*}
  E_{\mathcal{D}}(\btheta) =& \sum_{L \in~\mathrm{doses}} \sum_{t_j}
  \left(\hat{A}_{ij}(t_j;\,L) - A_{ij}(t_j;\, \btheta,\, L)\right)^2
  \\ &\quad + \left(\hat{E}_{ij} (t_j;\, L)- E_{ij}(t_j;\,\btheta,\,
  L) \right)^2.
                              \label{eq:sse}
\end{align*}
We seek the parameter set $\btheta^*$ that minimises the discrepancy
between the model and the data:
\begin{equation}
  \btheta^{*} = \underset{\btheta}{\rm{argmin}} \,
  E_{\mathcal{D}}(\btheta), \quad
  \text{subject to $\btheta^* \geq \bold{0}$.}
  \label{eq:min-prob}
\end{equation}

\begin{figure}[tp]
\centering
   \includegraphics[width=\textwidth]{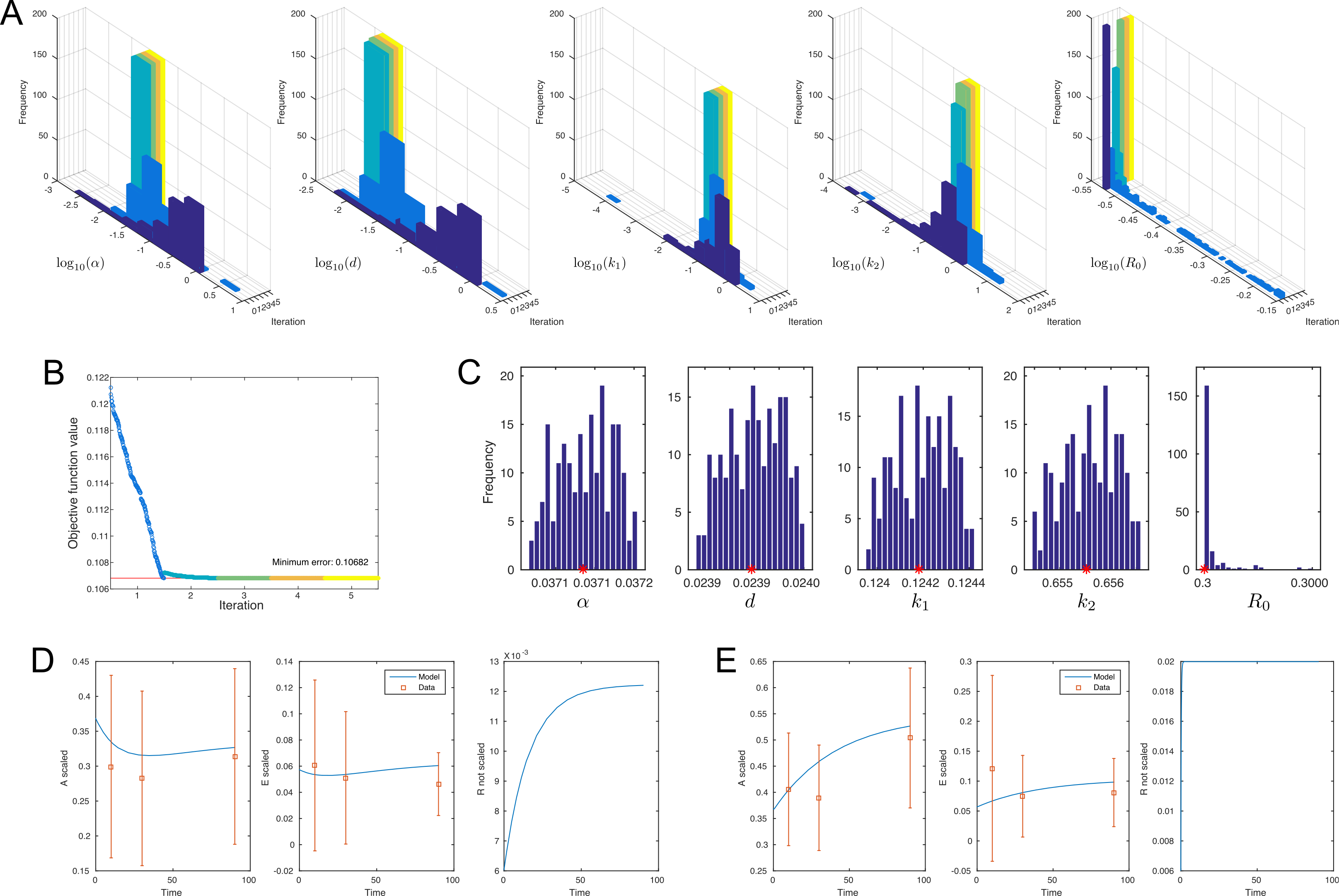}
  \caption{Example output of the Squeeze and Breathe parameter
    estimation algorithm. A: Sequence of posteriors for each parameter
    after each iteration. B: Convergence of the objective
    function. The plot shows the error of best parameter sets after
    each iteration. C: Final distribution of the parameters after
    convergence. The red star marks the value of the parameter with
    the lowest error. D and E: Time course of the model with the best
    parameters for one cluster. } \label{fig:hist}
\end{figure}

We find $\btheta^*$ using the Squeeze-and-Breathe (SB) evolutionary
optimisation algorithm~\cite{Beguerisse2012b}, available as code here {https://people.maths.ox.ac.uk/beguerisse/\#squeezecode}. Given an initial
estimate of the distribution of the parameter values (a `prior'). SB
generates a large number of parameter sets using Monte-Carlo
simulation. These points are used as the starting guess for a local
minimisation of $ E_{\mathcal{D}}(\btheta)$ using a derivative free
method such as Nelder-Mead~\cite{Nelder1965}. The local minima are
ranked and the best points are used to recompute the distribution of
the parameters (the `posterior'). These new posteriors are used as
priors in a new iteration of the algorithm, which again looks for new
local minima and keeps the best. This process goes on until the error
function converges. One of the key advantages of SB is that it can
handle situations where little is known about the parameter values by
efficiently exploring the parameter space, even venturing to regions
outside the original prior (for this reason the posteriors are not
true posteriors in the Bayesian sense). Figure~\ref{fig:hist} shows an
example for the type of output produced by SB.

\subsection{Model selection}
\begin{figure}[tp]
\centering
   \includegraphics[width=9.5cm]{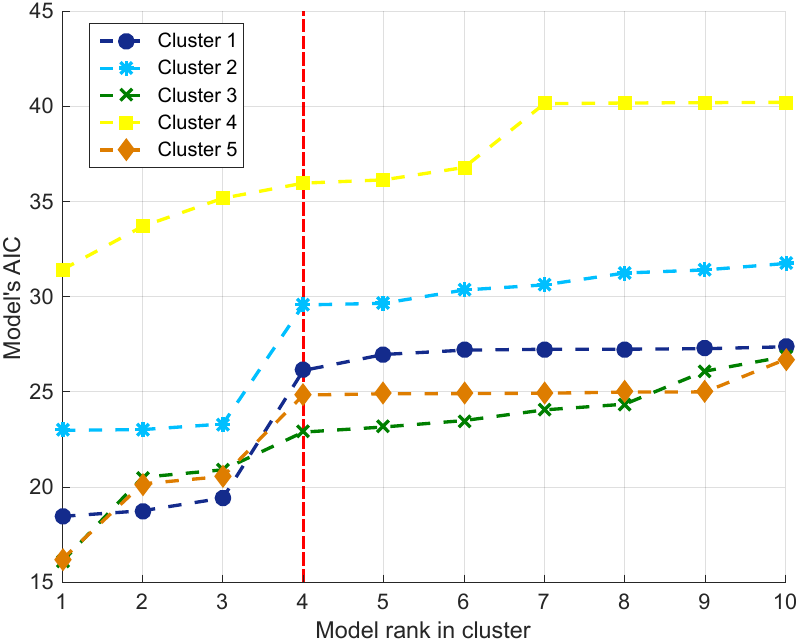}
  \caption{Values of the AIC for the models ranked first to tenth in
    each of the five clusers depicted in Fig.~4 of the Main Text. The
    dashed red line indicates the cutoff for the models in
    Fig.~4.} \label{fig:aic_decay}
\end{figure}

We perform model selection on the 40 models and 5 clusters using the
Akaike Information Criterion with a correction for finite sample size
(AICc)~\cite{Burnham2003}. For a given model $i$ with $p_i$
parameters,
$$AICc_i = n\ln(nRSS_i) + 2p_i+ \frac{2p_i(p_i+1)}{n-(p_i+1)},$$ where
$n$ is the number of observations and $RSS_i$ is the residual sum of
squares of the model. The AICc balances how well the model fits the
data with the complexity of the model (the number of parameter
values). The lowest AICc$_i$ is the preferable model.
Figure~\ref{fig:aic_decay} shows the AIC scores of the models for each
cluster, ranked from first (best) to tenth.

\bibliographystyle{pnas}

\end{document}